\documentclass[pra,superscriptaddress,twocolumn]{revtex4-1}

\usepackage{graphicx}% Include figure files
\usepackage{dcolumn}% Align table columns on decimal point
\usepackage{bm}% bold math
\usepackage{amsmath,amssymb}
\usepackage{amsfonts}
\usepackage{txfonts}
\usepackage{mathtools}
\usepackage{braket}
\usepackage{url}
\usepackage{multirow}
\usepackage{arydshln}
\usepackage{algorithm}
\usepackage{algorithmic}
\usepackage{colortbl}
\usepackage[colorlinks]{hyperref}
\usepackage{color}

\usepackage{ulem}

\begin{document}

%\preprint{APS/123-QED}

\title{Characterizing quantum pseudorandomness by machine learning}
%\thanks{A footnote to the article title}

\author{Masahiro Fujii}
 \affiliation{Department of Informatics, Graduate School of Integrated Science and Technology, Shizuoka University, 3-5-1 Johoku, Naka-ku, Hamamatsu 432-8011, Japan}
 
\author{Ryosuke Kutuzawa}
 \affiliation{Department of Informatics, Graduate School of Integrated Science and Technology, Shizuoka University, 3-5-1 Johoku, Naka-ku, Hamamatsu 432-8011, Japan}

\author{Yasunari Suzuki}
 \affiliation{NTT Computer and Data Science Laboratories, Musashino 180-8585, Japan}
 \affiliation{JST, PRESTO, 4-1-8 Honcho, Kawaguchi, Saitama, 332-0012, Japan}
 
\author{Yoshifumi Nakata}
 \affiliation{Photon Science Center, Graduate School of Engineering, The University of Tokyo, 7-3-1 Hongo, Bunkyo-ku, Tokyo 113-8656, Japan}
 \affiliation{JST, PRESTO, 4-1-8 Honcho, Kawaguchi, Saitama, 332-0012, Japan}
 
\author{Masaki Owari}
 \email{masakiowari@inf.shizuoka.ac.jp}
 \affiliation{Department of Informatics, Graduate School of Integrated Science and Technology, Shizuoka University, 3-5-1 Johoku, Naka-ku, Hamamatsu 432-8011, Japan}
% \affiliation{Faculity of Informatics, Shizuoka University, 3-5-1 Johoku, Naka-ku, Hamamatsu 432-8011, Japan}

\date{\today}% It is always \today, today,

\begin{abstract}
Random dynamics in isolated quantum systems is of practical use in quantum information and is of theoretical interest in fundamental physics. Despite a large number of theoretical studies, it has not been addressed how random dynamics can be verified from experimental data. In this paper, based on an information-theoretic formulation of random dynamics, i.e., unitary $t$-designs, we propose a method for verifying random dynamics from the data that is experimentally easy-to-access. More specifically, we use measurement probabilities estimated by a finite number of measurements of quantum states generated by a given random dynamics.
Based on a supervised learning method, we construct classifiers of random dynamics and show that the classifiers succeed to characterize random dynamics. 
We then apply the classifiers to the data set generated by local random circuits (LRCs), which are canonical quantum circuits with growing circuit complexity, and show that the classifiers successfully characterize the growing features.
We further apply the classifiers to noisy LRCs, showing the possibility of using them for verifying noisy quantum devices, and to monitored LRCs, indicating that the measurement-induced phase transition may possibly not be directly related to randomness.
\end{abstract}

\maketitle
\clearpage

\section{Introduction}
Random unitary dynamics in quantum systems has been widely studied in theoretical physics, applied mathematics, and quantum information science. Random unitary dynamics is typically formulated by a Haar random unitary, also known as the circular unitary ensemble (CUE)~\cite{M2014RM}, and characterizes the dynamics in fully chaotic quantum systems. 
Recently, an approximation of a Haar random unitary has been proposed in quantum information, which is called a \emph{unitary $t$-design}~\cite{AE2007, L2010} and is defined as an ensemble of unitaries that approximates a Haar random unitary up to the $t$th order. 

Despite its mathematical appearance, a unitary $t$-design is an extremely successful concept. In quantum information, it is used as resource of many tasks of quantum information processing, such as communication~\cite{D2005, DW2004, GPW2005, ADHW2009, DBWR2010, SDTR2013, HOW2005, HOW2007, NWY2021}, cryptography~\cite{AS2004, HLSW2004}, quantum algorithms~\cite{S2005, BH2013}, sensing~\cite{KRT2014,KL15,KZD2016,OAGKAL2016}, experimental verification of noisy quantum devices~\cite{EAZ2005,KLRetc2008,MGE2011,MGE2012}, and potentially quantum computational supremacy~\cite{B2018, G2019,BFNV2019}.
In theoretical physics, unitary $t$-designs have been explored so as to quantitatively characterize quantum chaos and have been spiking a large number of research topics, such as local thermalization in isolated quantum systems~\cite{PSW2006,dRARDV2011,dRHRW2016}, information paradox of quantum black holes~\cite{HP2007,SS2008,S2011,LSHOH2013,HQRY2016,RY2017, NWK2020, LFSLYYM2019}, and measurement-induced phase transitions~\cite{LCF2018,JYVL2020,BCA2020}.

Motivated by its importance in quantum information and theoretical physics, implementing a unitary $t$-design is of practical and theoretical interest. A unitary $t$-design can be efficiently realized by quantum circuits~\cite{HL2009TPE,BHH2016,HM2018,HMHEGR2020, NZOBSTHYZTTN2021} and by the dynamics of a generalized kicked Ising model~\cite{NHKW2017}, based on which unitary designs have been already realized in small systems~\cite{B2018, Lietal2019}.

Various methods have been proposed to characterize $t$-design \cite{B2018,RY2017,Scott_2008,LO1969, Almheirietal2013,Shenker2014,Shaffer_2014,CHM2014,YHGMC2017,Liu2018,BCHKP21,Iaconis2021}.
However, most of them include 
the use of a process tomography, which can reveal all the details of the dynamics at the cost of exponential computational resource. Since this way is far from tractable in a large system,
it is extremely hard to check whether experimentally-implemented dynamics is a unitary $t$-design, and it is of significant importance to look for tractable verification methods of unitary $t$-designs from experimentally-accessible data.

In recent years, machine learning has become a powerful tool for investigating physics in large systems. For instance, machine learning is shown to be useful in the reconstruction of quantum states, identification of phase transition, and modelling thermodynamic observables~\cite{PhysRevB.94.165134,doi:10.1126/science.aag2302,torlai_neural-network_2018,PhysRevX.8.021050,carrasquilla_reconstructing_2019,doi:10.1146/annurev-conmatphys-031119-050651}.
A machine learning approach also turned out to be useful for verification of $t$-designs~\cite{alves2020machine} based on the data set of diagnostics of quantum chaos~\cite{LO1969, Almheirietal2013}, namely, Out-of-Time-Ordered Correlators (OTOCs).
By visualizing OTOCs and using a convolutional neural network, it was succeeded to distinguish a Haar random unitary, the Pauli group, which is a unitary $1$-design, and a unitary $2$-design generated by certain quantum circuits, in $10$-qubit systems with accuracy over $99\%$. 
%This method is accurate, but it is experimentally difficult to measure OTOCs. Hence, experimentally more tractable methods for verifying $t$-designs are desired.

In this paper, we propose a new supervised learning approach for verifying unitary $t$-designs from an easy-to-access experimental data set, far easier than OTOCs that are known to be hard to measure in experiment. We particularly use measurement outcomes in the computational basis.
We construct classifiers by learning the measurement outcomes of quantum states generated by random Clifford unitaries (RCs), which is a canonical instance of unitary $3$-designs~\cite{Zhu2017, W2016}, and those by a Haar random unitary. The classifiers are shown to successfully distinguish the two with adequate accuracy, implying that verifying unitary $t$-designs from measurement outcomes in the computational basis is possible.
We emphasize that our approach is based on a set of a realistic number of measurement outcomes, which contains statistical errors caused by the finite number of measurements. Hence, unlike the previous method based on the ideal averages of OTOCs, our classifiers can be used to verify designs from actual experimental data with finite length.

We then check if the classifiers succeed to characterize the dynamics that is not used in the learning process. To this end, we apply the classifiers to the measurement outcomes of local random circuits (LRCs) with various depth. We use LRCs since they are canonical quantum circuits that exhibit growth of circuit complexity \cite{Haferkamp2022} in the sense of $t$-designs by varying the depth, i.e., deeper LRCs form unitary $t$-designs with larger $t$.
It turns out that, despite the fact that the measurement outcomes of LRCs have a data structure entirely different from RCs and a Haar random unitary, the classifiers are able to characterize the growth of complexity in LRCs.
This strongly suggests that our classifiers are useful for characterizing $t$-designs of any types, even if it substantially differs from the unitaries used at the learning stage.

To further demonstrate the applicability of our classifiers, we also apply them to the measurement outcomes obtained from noisy LRCs, which are a toy model of experimentally-realized LRCs, and to those from the monitored LRCs that are commonly used to study measurement-based phase transitions (MIPTs)~\cite{LCF2018,JYVL2020,BCA2020}. We show that the classifiers are useful to detect tiny noise in LRCs, opening the possibility for using the classifiers to verify noisy quantum devices, and that the MIPTs are not detectable by our classifiers, possibly indicating that MIPTs may not be transitions in the sense of $t$-designs.

This paper is organized as follows. We start with preliminaries in Sec.~\ref{sec of detail setting} and provide a summary of results in Sec.~\ref{s.o.r}.
After the organization of the rest is presented in Sec.~\ref{Sec:OrR}, we explain our methods in detail. In Sec.~\ref{S:FVdetail}, we explain how to generate the date for learning and classifiers. 
The learning methods are explained in detail in Sec.~\ref{creation and evaluation about the discriminator}. Classifications of LRCs are presented in Sec.~\ref{S:CLRCsRDCs}. In Sec.~\ref{sec of application}, we apply our classifiers to the noisy LRCs and the monitored LRCs. After we conclude the paper in Sec.~\ref{sec of conclusion}, technical details are given in Appendices.

\section{Preliminaries} \label{sec of detail setting}
We provide definitions and basic properties of unitary $t$-designs, and quantum circuits that we will use in our analysis.

\subsection{A Haar random unitary and unitary $t$-designs}\label{subsec Haar measure}
A Haar random unitary is defined as a unitary drawn uniformly at random with respect to the uniform measure, i.e. the Haar measure, on the unitary group. 
In this paper, we denote a Haar random unitary by ${\sf Haar}$.
Despite its abstract formulation, ${\sf Haar}$ finds a huge number of applications in quantum information~\cite{D2005, DW2004, GPW2005, ADHW2009, DBWR2010, SDTR2013, HOW2005, HOW2007, NWY2021,AS2004, HLSW2004, S2005, BH2013, KRT2014,KL15,KZD2016,OAGKAL2016, EAZ2005,KLRetc2008,MGE2011,MGE2012, B2018, G2019,BFNV2019}, as useful resource for information processing, and in theoretical physics~\cite{PSW2006,dRARDV2011,dRHRW2016, HP2007,SS2008,S2011,LSHOH2013,HQRY2016,RY2017, NWK2020, LFSLYYM2019}, as typical dynamics in fully chaotic quantum systems.

Despite its usefulness, ${\sf Haar}$ cannot be realized in realistic timescale when the system size is large~\cite{emerson2003pseudo,PQSV11}.  A unitary $t$-design was introduced to circumvent the difficulty.
Let $t$ be a positive integer and ${\sf W}$ be an ensemble of unitaries in $U(d)$. We define a quantum channel $G^{(t)}[{\sf W}]$ by
\begin{align} \label{def of channnel}
  G^{(t)}[{\sf W}](\rho) &:= \mathbb{E}_{U \sim {\sf W}} \left[ U^{\otimes t} \rho U^{\dagger \otimes t} \right],
\end{align}
where $\mathbb{E}_{U \sim {\sf W}}[f(U)]$ is an average of $f(U)$ over the ensemble ${\sf W}$.
An ensemble ${\sf U}_t$ of unitaries is called a unitary $t$-design if 
\begin{equation} \label{def of t-design}
  G^{(t)}[{\sf Haar}] = G^{(t)}[{\sf U}_t].
\end{equation}
By definition, ${\sf Haar}$ is a unitary $t$-design for any $t$. Hereafter, we refer to a unitary $t$-design as a $t$-design for short.

When Eq.~\eqref{def of t-design} is satisfied only approximately, it is called an $\epsilon$-approximate $t$-design ${\sf U}_{t, \epsilon}$, that is,
\begin{equation}
\bigl \|  G^{(t)}[{\sf Haar}]- G^{(t)}[{\sf U}_{t, \epsilon}]  \bigr \|_{\diamond} \leq \epsilon,
\end{equation}
where $|| \cdot ||_{\diamond}$ is the diamond norm~\cite{KSV2002}, for $\epsilon \geq 0$.

Note that a unitary $t$-design is not by any means unique. For instance, when an $\epsilon$-approximate $t$-design ${\sf U}_{t, \epsilon}$ is shifted by applying a fixed unitary, it is also an $\epsilon$-approximate $t$-design, which is not identical with the original one in general. Note also that a $t$-design is always a $(t-1)$-design, which follows from the definition.

Recalling that ${\sf Haar}$ is equivalent to the CUE that models typical complex dynamics in quantum chaotic systems, $t$-designs can be thought of as approximations of quantum chaotic dynamics. From this viewpoint, the degree $t$ of $t$-designs quantifies how chaotic the dynamics is. In other words, $t$ can be regarded as an indicator of complexity of random dynamics. This is the idea behind a number of studies in theoretical physics~\cite{HP2007,SS2008,S2011,LSHOH2013,HQRY2016,RY2017, NWK2020, LFSLYYM2019}.

\subsection{Random quantum circuits realizing $t$-designs} \label{subsec Random quantum circuits}
A $t$-design can be generated by quantum circuits for all $t$ even exactly~\cite{BNOZ2020, NZOBSTHYZTTN2021}. There also exist a couple of families of random quantum circuits that efficiently and approximately implement approximate $t$-designs~\cite{HL2009TPE,BHH2016,HM2018,HMHEGR2020, NZOBSTHYZTTN2021}. We here overview three classes of random unitaries that form $t$-designs: Random Clifford unitaries (RCs), Local Random Circuits (LRCs) and Random Diagonal Circuits (RDCs). 

The RCs are random unitaries sampled from the uniform distribution on the Clifford group. Let ${\sf P}_n = {\sf P}_{1}^{\otimes n}$ be the Pauli group acting on $n$ qubits, where ${\sf P}_{1} =  \langle \sqrt{-1}I, X, Y, Z \rangle$ with
\begin{align}
& I=\begin{bmatrix} 1 & 0 \\ 0 & 1 \\ \end{bmatrix}, X=\begin{bmatrix} 0 & 1 \\ 1 & 0 \\ \end{bmatrix}, Y=\begin{bmatrix} 0 & -\sqrt{-1} \\ \sqrt{-1} & 0 \\ \end{bmatrix}, Z=\begin{bmatrix} 1 & 0 \\ 0 & -1 \\ \end{bmatrix}.  \nonumber
\end{align}
The basis $\{\ket{0}, \ket{1}\}$ of the Pauli-$Z$ operator is called the computational basis. Unlike the convention, we define the computational basis such that
\begin{equation}
Z = - \ket{0}\!\bra{0} + \ket{1}\!\bra{1}.
\end{equation}
The $n$-qubit Clifford group ${\sf C}_n$ is given by
\begin{equation}
\label{DEFC}
{\sf C}_n := \{U \ | \ UPU^{\dagger} \in {\sf P}_{n}, \forall P \in {\sf P}_{n}\}.
\end{equation}
When a unitary is drawn from ${\sf C}_n$ uniformly at random, we refer to is as a random Clifford unitary and denote it by ${\sf RC}$. It is known that ${\sf RC}$ is an exact $3$-design~\cite{Zhu2017, W2016} (and hence an exact $2$-design) in \emph{qubit}-systems, although it is not even a $2$-design in  general \emph{qudit}-systems~\cite{GSW2021}.

An LRC is a random quantum circuit consisting of multiple layers~\cite{BHH2016}. In odd-numbered layers, random two-qubit gates, independently sampled from the Haar measure on $U(4)$, are applied on the $(2i-1)$-th and $2i$-th qubits for all $i$. Similarly, in even-numbered layers, random two-qubit gates are applied to $2i$-th and $(2i+1)$-th qubits for all $i$.
An LRC defines a random unitary ${\sf LRC}(D)$, which is known to be an $\epsilon$-approximate $t$-design if the depth $D$ (that is, the number of layers) of the circuit satisfies~\cite{BHH2016}
\begin{equation} \label{LRC depth and design}
  D \geq c t^{9}\left (n \cdot t + \log_{2}\frac{1}{\epsilon}\right),
\end{equation}
where $c$ is a constant. Hence, a deeper LRC achieves a $t$-design with larger $t$. 
Recalling that the degree $t$ of a $t$-design indicates how complex the dynamics is, ${\sf LRC}(D)$ exhibits growth of circuit complexity in the sense of $t$-designs as the depth increases.

An RDC is a random quantum circuit that iterates a random diagonal circuit $M$, composed of $(n^2-n)/2$ random two-qubit gates diagonal in the Pauli-$Z$ basis, and Hadamard gates $H^{\otimes n}$~\cite{NHKW2017}. When the number of iterations is $I$, we denote the corresponding random unitary by ${\sf RDC}(I)$. Note that the depth is $\approx I n/2$.
It was shown in~\cite{NHKW2017} that, for $t=o(\sqrt{n})$, ${\sf RDC}(I)$ is an $\epsilon$-approximate unitary $t$-design if
\begin{equation} \label{RDC depth and design}
  I \geq \frac{1}{n - 2 \log_2 t!}\bigl(t n + \log_{2} 1/\epsilon \bigr).
\end{equation}
Thus, by using RDCs, a $t$-design is implemented by a circuit with depth $tn/2$.

Although both ${\sf LRC}(D)$ and ${\sf RDC}(I)$ were shown to be a $t$-design if the depth is $t^{10} n$ and $tn/2$, respectively, and ${\sf RDC}(I)$ may seem much more efficient than ${\sf LRC}(D)$, it is conjectured that ${\sf LRC}(D)$ with $D= O(tn)$ is a $t$-design. This is partly proven for small $t$'s~\cite{HJ2019}, but it remains open for general $t$.

%We also estimated an upperbound of $\frac{F_\mu^{(t)}-F_{\mu_H}^{(t)}}{t!}$ for a RDC for each $t$ and depth in a $7$-qubit system. The result is given in FIG.  \ref{graph of epsilon of RDC}.
%FIG.  \ref{graph of epsilon of RDC} leads that a RDC with depth greater than $4$ is an $\epsilon$-approximated $t$-design with $t\leq5$, $\epsilon<0.05$.

\subsection{Supervised machine learning}

Supervised machine learning is a framework of machine learning. In this framework, our task aims to train a prediction model that estimates a label from its features using the provided pairs of features and labels. To this end, we prepare a parameterized estimation model and fit the parameters so as to minimize the discrepancy between the correct labels from estimated labels. The measure of discrepancy is called loss, and this phase is called training.

Since the trained prediction model may be over-fitted to the training data, the above loss score would be overestimated. Thus, we typically split the given data set into three parts, training, validation, and test dataset, in advance, and one of them is used for training and another is for validating the performance of the prediction model. By monitoring the validation score, we stop the training when the validation score is reduced to a reasonable value. 

After the prediction model is trained by minimizing the validation scores, we evaluate the final performance of the prediction models with the test dataset. Note that in our case, we sometimes want to let the prediction model classify the data sampled from a different population from the trained dataset. In this case, we provide another dataset in the test phase and treat its loss as the prediction performance.

\section{Summary of results} \label{s.o.r}

We summarize our results about supervised learning of random dynamics.
After explaining two criteria for classifiers in Subsec.~\ref{SS:TCC}, we construct several classifies based on different learning methods in Subsec.~\ref{SS:TrainingSum}.
In Subsecs.~\ref{SS:LRCSum} and~\ref{SS:LRCvarSum}, we provide brief summaries of applications of the classifiers. In the former subsection, we show that the classifiers can be used to detect the growth of circuit complexity in LRCs. In the latter subsection, we apply the classifiers to noisy and monitored LRCs.

\subsection{Two criteria of classifiers} \label{SS:TCC}

The main goal of this paper is to construct a classifier that can characterize random dynamics in the sense of $t$-designs from an experimentally-accessible data set. The classifier should satisfy the following two criteria:
\begin{enumerate}
\item the data set is easy-to-access in experiments,
\item the classifier characterizes a given dynamics in terms of $t$-designs and should not be too dependent on the details of the dynamics.
\end{enumerate}

The first one is of significant importance for the method to be useful for practical experiments. Although there is ambiguity of `easy-to-access in experiments', we here assume that it is possible to separately measure each qubit in the computational basis. For simplicity of analysis, we assume that the measurement is noiseless. 
Note, however, that measurement probabilities themselves are not directly accessible since computing them requires infinitely many repetitions of measurements, which is not tractable.

To be precise, let $N_q$ be the number of qubits in the system and ${\sf U} = \{U_i\}_i$ be an ensemble of implemented unitaries, where $i$ can be a continuous variable and a prior probability distribution over $i$ may not be uniform. We sample $N_u$ unitaries from ${\sf U}$. For each sampled unitary $U_i$, we generate a state 
\begin{equation} \label{eq of quantum circuit}
  \ket{\psi_{out}^{(i)}} = U_i  P  \ket{0}^{\otimes N_q},
\end{equation}
where $P$ represents pre-processing and is applied if necessary. The initial state is set to a computational-basis state.
We then measure each qubit of the output state in the computational basis. 
After the measurement, a bit string of length $N_q$ is obtained. 
We perform this procedure $N_s$ times for the same unitary $U_i$, from which we estimate the probability to obtain each measurement outcome. Here, $N_s$ is chosen to be no greater than $10000$, which we believe is a realistic value for an experiment using existing quantum devices.
From this procedure, we obtain estimated values $J_{est}(h_1,\dots,h_k,k'|N_u,N_s)$ of the $k'$th moment $J(h_1, \dots, h_k, k')$ of $k$-qubit correlation among the $h_1$th, $\dots$, $h_k$th qubits, where
\begin{equation}
J(h_1, \dots, h_k, k') = \mathbb{E}_{U_i \sim {\sf U}}\biggl[\bra{\psi_{out}^{(i)}} Z_{h_1} \otimes \dots \otimes Z_{h_k} \ket{\psi_{out}^{(i)}}^{k'} \biggr].
\end{equation}
Here, we use the notation that the identity operator is implicit.
See Eqs.~\eqref{Eq:111} and~\eqref{eq_bitcorr_2 est2} for the formal expression of the estimators $J_{est}(h_1,\dots,h_k,k'|N_u,N_s)$.
Note that
\begin{equation}
    \lim_{N_u, N_s \rightarrow \infty}J_{est}(h_1,\dots,h_k,k'|N_u,N_s) = J(h_1, \dots, h_k, k'). \label{Eq:infinite}
\end{equation}
It is naively expected that $k'$-designs are likely to be distinguished from higher-designs if one is allowed to access the values of $J(h_1, \dots, h_k, k')$, since they are $k'$-th moments of the unitary. However, it is not clear if we can do so from $J_{est}(h_1,\dots,h_k,k'|N_u,N_s)$ that contains statistical errors due to the finite number of measurements.

A set of estimated values $\{ J_{est}(h_1,\dots,h_k,k'|N_u,N_s) \}$ for all choices of $(h_1, \dots, h_k)$ and for all $k \in \{1,\dots, N_q\}$ defines a data set for the supervised learning.
Each is called a \emph{feature vector}. Note that only quantum part is to generate $\ket{\psi_{out}^{(i)}}$ and to measure it in the computational basis. See Subsec.~\ref{SS:GDS} for more details. 

The second criterion is also important since we are interested in the properties of a given dynamics with respect to $t$-designs rather than its precise description. Hence, the classifier should not be too dependent on the details of the dynamics. 
It is highly non-trivial if this criterion can be satisfied when we aim to classify $t$-designs by supervised learning methods because there are infinitely many distinct $t$-designs for each $t$ but it is impossible to learn all of them. Instead, what one can practically try is to pick up a specific $t$-design for a fixed $t$ and use it for learning the properties of $t$-designs. In this case, the classifier may learn the properties specific only to that ensemble, which we do not wish.

In our approach, to check the second criterion, we apply obtained classifiers to the data sets generated by $t$-designs different from those used at the learning stage. If the classification succeeds for different types of $t$-designs as well, we can reasonably conclude that the classifier correctly captures the properties of $t$-designs.

\subsection{Training methods and evaluating performance} \label{SS:TrainingSum}

Using the set of feature vectors, i.e., the data set, we train classifiers for $t$-designs. We use in the training stage the data set generated by ${\sf RC}$, where the pre-processing $P$ is chosen to be a \emph{fixed} unitary drawn uniformly at random from ${\sf Haar}$, and that by ${\sf Haar}$, where $P = I$.
They are generated by classical simulations. 
The non-trivial pre-processing $P$ in the training process of ${\sf RC}$ is important to keep the classifiers away from learning the properties very specific to ${\sf RC}$. See Appendix~\ref{Fixed_basis} for the details.

Our purpose is to construct classifiers which distinguish $3$-designs from higher-designs if learning succeeds.
We expect that the classifier trained by the data set with $k'=4$ satisfies this condition because it uses only $4$th moments of qubit correlations.
Note that, even though ${\sf RC}$ is a $3$-design but not a higher-design, and ${\sf Haar}$ is a $t$-design for all $t$, this does not immediately mean that any classifier trained by this data satisfies the above property because of the nature of supervised learning. 
It is also not clear if the classifier with $k'=4$ is superior to the classifiers with higher $k'$ in advance, letting us to try other cases as well in later sections.
Here, we present the results for $10$ qubits and the moment $k' = 4$. Other cases are investigated in later sections.

By classically simulating ${\sf RC}$ and ${\sf Haar}$, we generate data sets, namely, feature vectors consisting of $J_{est}(h_1,\dots,h_k,k'|N_u,N_s)$, for training, validation, and testing. From the data sets for training and validation, we construct classifiers, which output either ${\sf RC}$ or ${\sf Haar}$, using different learning algorithms, i.e., linear regression (LR), linear support vector machine (LSVM), support vector machine (SVM), random forest (RF), and feed forward neural network (NN). See Appendices~\ref{ML_algos} and~\ref{append_envs} for the setting of learning parameters and for our computational resource, respectively.
To avoid the classifiers to learn the properties that are very specific to the training data set, we shuffle the training and validation data sets and construct $10$ classifiers for each algorithm. The performance of each algorithm is determined by the average over the $10$ classifiers. See Subsec.~\ref{SS:TM} for more details of the training methods.

We then use the test data to evaluate the performance of ten classifiers of each algorithm. The average success probability of classification, where the average is taken over ten classifiers, is given in TABLE.~\ref{table_10q_ASP}. We clearly observe that only the classifiers based on SVM and NN, which we refer to as SVM and NN classifiers, respectively, succeed in classifying the data. 
The fact that only SVM and NN successfully learn random dynamics of RCs and a Haar random unitary is reasonably understood as follows. We can suppose $d$-dimensional feature vectors as elements in the $d$-dimensional vector space called a feature space. The classifiers set a decision boundary, which divides the feature space into two regions, and the data points are classified into two classes according to the region they belong to. LR and LSVM are linear classifiers, which put a hyperplane as the decision boundary. Thus, they cannot discriminate between data points that cannot be separated linearly.
In contrast, SVM and NN can perform a non-linear separation using a kernel function and non-linear activation, respectively. Since our classification task is expected to demand complicated, it seems consistent that non-linear classification is required for achieving high accuracy. This is also consistent with the principal component analysis in Sec.\,\ref{sec of application}. 
RF is also a classifier that can achieve non-linear classification with ensemble learning and shows the perfect classification for the training datasets. However, it does not work for validation and test datasets and cannot be used for the unknown datasets. See Sec.\,\ref{SS:TM} for details on this point.

%\red{The SVM and NN classifiers also have different advantages..... We also note that a learning algorithm using NN is much faster than that using SVM in general.}

The fact that LR and LSVM hardly succeed in learning the properties of unitary designs highlights the usefulness of machine learning approaches in our task. This is because the failures of the two indicate that linear functions of the moments of correlations are not sufficient for the characterization, namely, we need non-linear functions to characterize $t$-designs. Finding an appropriate non-linear function is in general intractable, but sophisticated machine learning techniques, such as NN or (non-linear) SVM, provide powerful tools in search of appropriate non-linear functions, as we have clearly demonstrated.
Note that the necessity of non-linear functions for the characterization should be closely related to the statistical error caused by the finite number of measurement outcomes. If infinite number of measurements is allowed in the analysis, linear functions may possibly succeed the characterization.

The accuracy of the classification can be improved by using larger data sets of $J_{est}(h_1,\dots,h_k,k'|N_u,N_s)$. In fact, it is shown in a $7$-qubit system that, by increasing the size of each data set, the accuracy can be more than $85$\%, which is naturally expected from Eq.~\eqref{Eq:infinite}. See Sec.\ref{creation and evaluation about the discriminator} for the details.

From this result, we conclude that, by using the SVM and NN algorithms, it is possible to learn properties of ${\sf RC}$ and ${\sf Haar}$ from the data sets that are experimentally easy-to-access. 

Here, we remark on the prprocessing unitary ($P$), which is a fixed unitary sampled with respect to the Haar measure. It is only for making the training method more effective. After the classifiers are sufficiently trained, $P$ is not necessarily applied in experiments. We also note that the training data is generated by classical simulation, and the generation of $P$ is not efficient in the sense of computational complexity since, although it is sampled only once, it is chosen from the Haar measure. However, in small systems, it is easy and little time-consuming to generate the Haar measure by using existing packages of simulations. For this reason, we have chosen the fixed unitary $P$ from the Haar measure. If necessary, $P$ can be replaced with local random circuits with sufficient depth, which will not cause any substantial changes on the results. 

\setlength{\tabcolsep}{6pt} % Default value: 6pt
\renewcommand{\arraystretch}{1.2} % Default value: 
\begin{table}[tb!]
  \centering
  \caption{Average success probability of classification for training, validation and test data sets. The average and standard deviation are taken over $10$ classifiers constructed from the same algorithm.}
  \begin{tabular}{c c c c}
    \hline
    Algorithms & training & validation & test \\
    \hline  \hline
    LR   & 0.592$\pm$0.002 & 0.500$\pm$0.004 & 0.501$\pm$0.002 \\
    LSVM & 0.592$\pm$0.002 & 0.502$\pm$0.003 & 0.501$\pm$0.004 \\
    SVM  & 0.965$\pm$0.002 & 0.601$\pm$0.003 & 0.605$\pm$0.002 \\
    RF   & 1.000$\pm$0.000 & 0.516$\pm$0.003 & 0.518$\pm$0.004 \\
    NN   & 0.966$\pm$0.020 & 0.672$\pm$0.004 & 0.675$\pm$0.003 \\
    \hline
    \label{table_10q_ASP}
  \end{tabular}
\end{table}

\subsection{Classifying LRCs} \label{SS:LRCSum}

\begin{figure}[t!]
  \centering
  \includegraphics[width=0.45\textwidth]{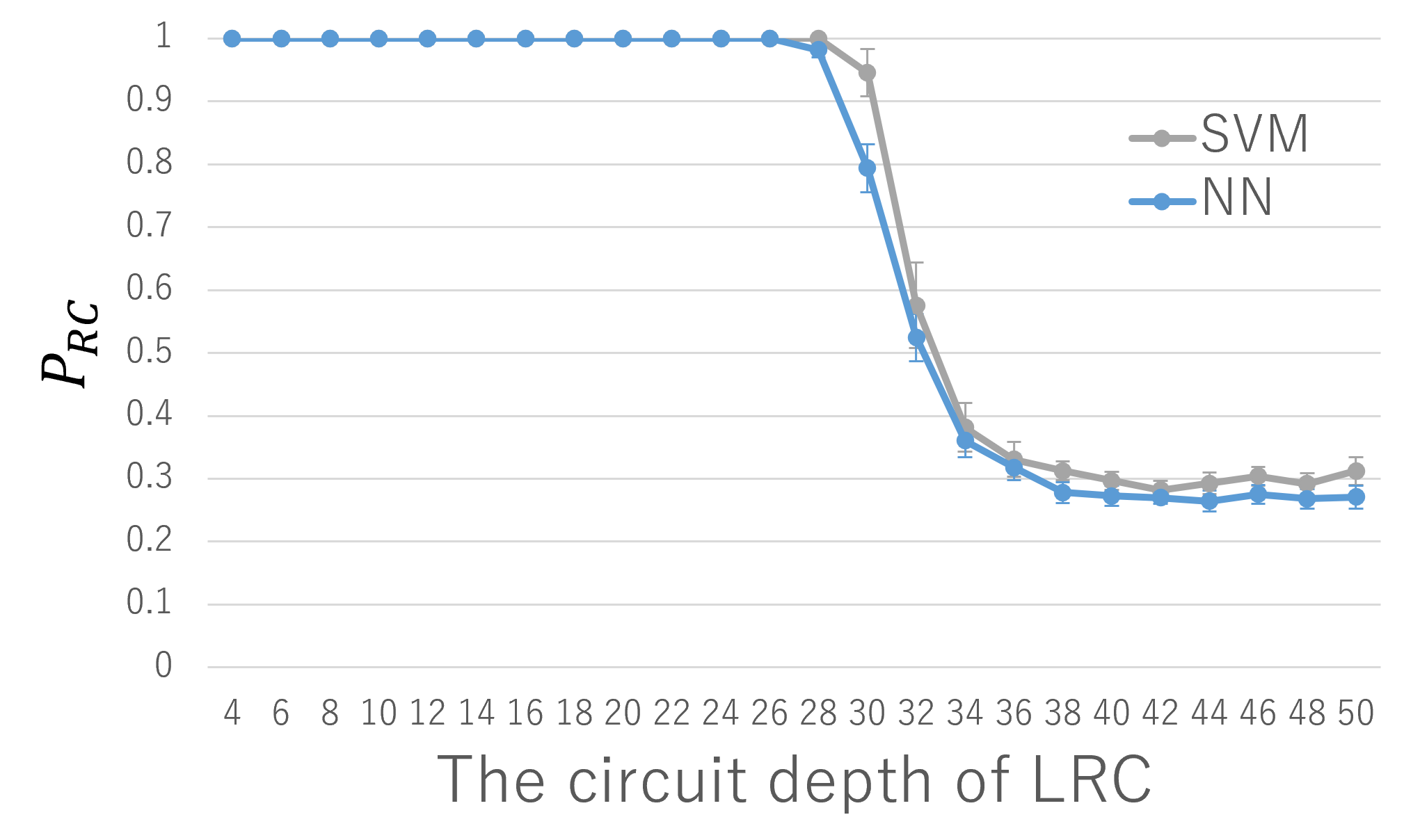}
  \caption{Classification results of the NN classifier with $k'=4$ for the data set generated by ${\sf LRC}(D)$ of $10$ qubits. The vertical axis is the probability $P_{RC}$ that the classifier outputs ${\sf RC}$, which defined by Eq.(\ref{eq def P RC}), and the horizontal axis is the depth $D$ of ${\sf LRC}(D)$. The dot is the average and the error bar is the standard deviation over $10$ distinct classifiers.}
  \label{pic_10q_5algs}
\end{figure}

We next apply the SVM and NN classifier, trained by the data set of ${\sf RC}$ and ${\sf Haar}$, to classifying other types of random unitaries, and check the second criterion posed in Subsec.~\ref{SS:TCC}, i.e., whether they succeed to classify the dynamics that substantially differs from those used in the training stage.

To this end, we apply the classifiers to the data set generated by ${\sf LRC}(D)$. As explained in Subsec.~\ref{subsec Random quantum circuits}, ${\sf LRC}(D)$ is a canonical quantum circuit that exhibits growth of circuit complexity in terms of $t$-designs by varying the depth. Hence, we expect that the classifiers succeed to characterize the growth.
Specifically, from the fact that ${\sf RC}$ is a $3$-design but not higher-design, and the classifier with $k'=4$ is defined by $4$th moment, we hope that the classifier with $k'=4$ detects the change of ${\sf LRC}(D)$ from a $3$- to a higher-design.

By classically simulating ${\sf LRC}(D)$ of 10 qubits, where we set $P=I$, we generate a data set of estimated values of $J_{est}(h_1,\dots,h_k,k'|N_u,N_s)$ for various depth. Each data set consists of $1000$ feature vectors. By inputting the data to the SVM and NN classifiers, we obtain an output of either ${\sf RC}$ or ${\sf Haar}$ for each feature vector. By repeating this for every feature vector, we have the probability $P_{RC}$ that the output of the classifier is ${\sf RC}$, which is formally given as follows. For any input feature vector $x$, a classifier $f$ outputs $f(x) \in \{ {\sf RC}, {\sf Haar} \}$.
Then, $P_{RC}$ is defined by 
\begin{equation}\label{eq def P RC}
 P_{RC}   := \frac{\left |\left \{x\in O_{\sf LRC}(D)\ | \ f(x)={\sf RC} \right \} \right|}{\left| O_{\sf LRC}(D) \right |},
\end{equation}
where $O_{\sf LRC}(D)$ is a set of feature vectors generated by ${\sf LRC}(D)$. 

Note that ${\sf LRC}(D)$ with finite $D$ is, by any means, neither ${\sf RC}$ nor ${\sf Haar}$. Hence, the output of the classifier, ${\sf RC}$ or ${\sf Haar}$, should not be literally taken. 
Similarly, $P_{RC}$ should be interpreted as the probability that, although ${\sf LRC}(D)$ differs from ${\sf RC}$, the corresponding ${\sf LRC}(D)$ is as random as, but not more random than, ${\sf RC}$.

The probability $P_{RC}$ for the SVM and NN classifiers to output ${\sf RC}$ is plotted in FIG.  \ref{pic_10q_5algs} as functions of $D$. We first observe that the outputs for the SVM and NN classifiers are similar to each other, which indicates that both work well for characterizing ${\sf LRC}(D)$.
It is also observed that $P_{RC}$ shows a transition around the depth between $28$ and $36$, namely, ${\sf LRC}(D)$ with small $D$ are classified to be ${\sf RC}$ and ${\sf LRC}(D)$ with $D \geq 36$ are likely to be classified as ${\sf Haar}$. Hence, the classifies succeed in capturing the growth of complexity in ${\sf LRC}(D)$ to a certain extent.

\setlength{\tabcolsep}{6pt} % Default value: 6pt
\renewcommand{\arraystretch}{1.2} % Default value: 
\begin{table}[tb!]
  \centering
  \caption{Average success probabilities of classification for training, validation and test data sets with various moments $k'$.}
  \begin{tabular}{c c c c}
    \hline
    order of $k'$ & training & validation & test \\
    \hline  \hline
    4  & 0.966$\pm$0.020 & 0.672$\pm$0.004 & 0.675$\pm$0.003 \\
    6  & 0.908$\pm$0.010 & 0.774$\pm$0.003 & 0.774$\pm$0.002 \\
    8  & 0.920$\pm$0.012 & 0.790$\pm$0.003 & 0.791$\pm$0.002 \\
    10 & 0.923$\pm$0.014 & 0.780$\pm$0.003 & 0.784$\pm$0.002 \\
    \hline
    \label{table_10q_ASP_evenk'}
  \end{tabular}
\end{table}

\begin{figure}[tb!]
  \centering
  \includegraphics[width=9cm]{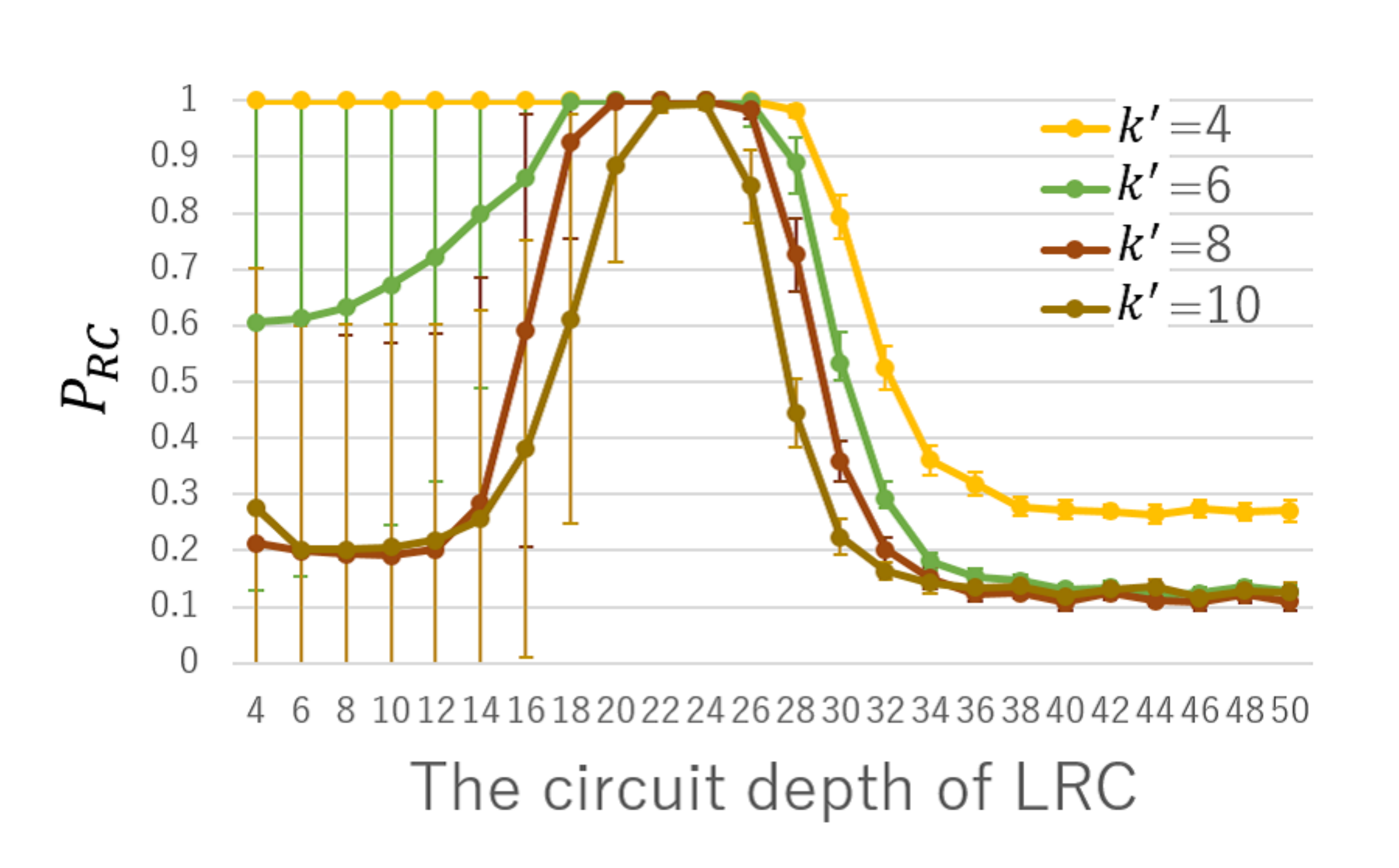}
  \caption{Classification results of the NN classifier with various $k'$ for the data generated by $10$-qubit ${\sf LRC}(D)$. The vertical axis is the probability $P_{RC}$ that the classifier outputs ${\sf RC}$, which defined by Eq.(\ref{eq def P RC}).}
  \label{pic_kp4_6_8_10_nn}
\end{figure}

It is, unfortunately, unclear if the change of complexity captured by the classifiers exactly correspond to the point from a $3$-design to a higher-design since a sufficient condition for ${\sf LRC}(D)$ to be a $t$-design (see Eq.~\eqref{LRC depth and design}) is unlikely to be tight.
However, we expect that exact changing points can be captured by the NN classifier since the NN classifier correctly characterizes the changing point from a $3$- to a higher-design in the case of ${\sf RDC}(I)$, where the changing point is analytically known. See~\ref{S:CLRCsRDCs} for the details. 

We also conjecture that the NN classifier is able to detect the exact depth, at which the change of the degree of designs, by taking the large limit of $k'$
%\Erase{, though it is theoretically not clear why the classifiers constructed with large $k'$ have such property, and further, the classifiers constructed with larger $k'$  are less stable against numerical precision.
To clarify this, we use the NN classifiers constructed from data sets $J_{est}(h_1,\dots,h_k,k'|N_u,N_s)$ with various moments $k'$. The result is given in FIG.~\ref{pic_kp4_6_8_10_nn} (see also TABLE.~\ref{table_10q_ASP_evenk'} for the performance of the classifiers). 

From FIG.~\ref{pic_kp4_6_8_10_nn}, we observe two features. First, the characteristic point gets smaller as the moment $k'$ increases. Second, the standard deviations of $P_{RC}$ for depth less than $\approx 20$ are so large that they cover most of the range. 
The second feature is reasonable since the classifier is trained by the data sets generated by ${\sf RC}$ and ${\sf Haar}$, both of which are $3$-designs. Hence, if a given dynamics is strictly less than $3$-designs, which is the case for ${\sf LRC}(D)$ with small $D$, the classifier shall return nearly a random output, resulting in a large standard deviation. For this reason, we should consider a large standard deviation as a sign that the dynamics is much less random than a $3$-design.
In FIG.~\ref{pic_kp4_6_8_10_nn}, the standard deviation is small when $D \geq 22$, indicating not only that the result is reliable for such $D$ but also that ${\sf LRC}(D)$ for such $D$ is at least a $3$-design.
This, together with the fact that the transition depth seems to approach $D = 22$ for larger $k'$, it is reasonable to guess that the depth, at which ${\sf LRC}(D)$ changes from a $3$-design to a higher-design, is around $D$ slightly more than $22$, such as $D$ between $24$ and $28$. 
Our conjecture is that this value will be more accurate by increasing $k'$,
%Here, we note that this result is unexpected, since  the classifier with $k'$ is defined based on $k'$th moments of qubit correlations. Thus, we expected that 
which may be counter-intuitive since it is natural to expect that the classifier with $k'$ is suitable for classifying $k'-1$ design and $t$-design with $t \ge k'$.

To summarize, we conclude that the NN classifier is capable for capturing the growth of circuit complexity in ${\sf LRC}(D)$ in the sense of $3$-designs. Since designs are approximate dynamics of quantum chaos, this can be rephrased as that the NN classifier can be used as quantitative diagnostics for complex chaotic dynamics.

Before we move on, we compare our method with the supervised learning approach based on OTOCs~\cite{alves2020machine}. See also Appendix~\ref{sec of add. exp. of alves's method}. 
The approach achieves $99\%$ accuracy, which is much better than our method with the accuracy less than $80\%$. Hence, if the values of OTOCs are available in experiment, the OTOC-based approach is likely to exceed the performance of our method. However, there are three caveats about practical uses of the OTOC-based approach. First, OTOCs are extremely hard to measure in experiment since they require the time-reversal dynamics as well, whose implementation is difficult except in the fully controllable systems such as NMRs or superconducting circuits~\cite{Li2017measuring,mi2021information}. Second, the OTOC-based approach~\cite{alves2020machine} uses the ideal values of OTOCs and does not account for the statistical error. The effect of statistical errors due to the finite number of measurements has not been clarified.
Finally, as shown in Appendix~\ref{sec of add. exp. of alves's method}, we reexamined the method and found that the OTOC-based approach does not seem to satisfy the second criteria in the subsection \ref{SS:TCC}: it is likely that it simultaneously learns the features not directly related to $t$-designs, which makes the classifier applicable only to the dynamics similar to those used in the training stage.
This indicates that the OTOC-based approach may work only when we have prior knowledge about the dynamics.

These three points indicate that our approach and the OTOC-based approach are in different use. Our classifiers are based on easy-to-access data and seem to work well even for the dynamics substantially different from those used in the training stage, but the result is with less accuracy. In contrast, the OTOC-based approach is based on hard-to-access data and seem to work only for the dynamics used in the training stage, but results in high accuracy.

\subsection{Characterizing various LRCs} \label{SS:LRCvarSum}

To demonstrate applicability of the NN classifier, we next apply it to LRCs with depolarizing noise, as a toy model of practical situations, and to the monitored LRCs, which shows MIPTs~\cite{LCF2018,JYVL2020,BCA2020}.
In the following, we set the preprocessing $P$ by $I$.

\subsubsection{LRCs with depolarizing noise}

\begin{figure}[tb!]
  \centering
  \includegraphics[width=0.35 \textwidth]{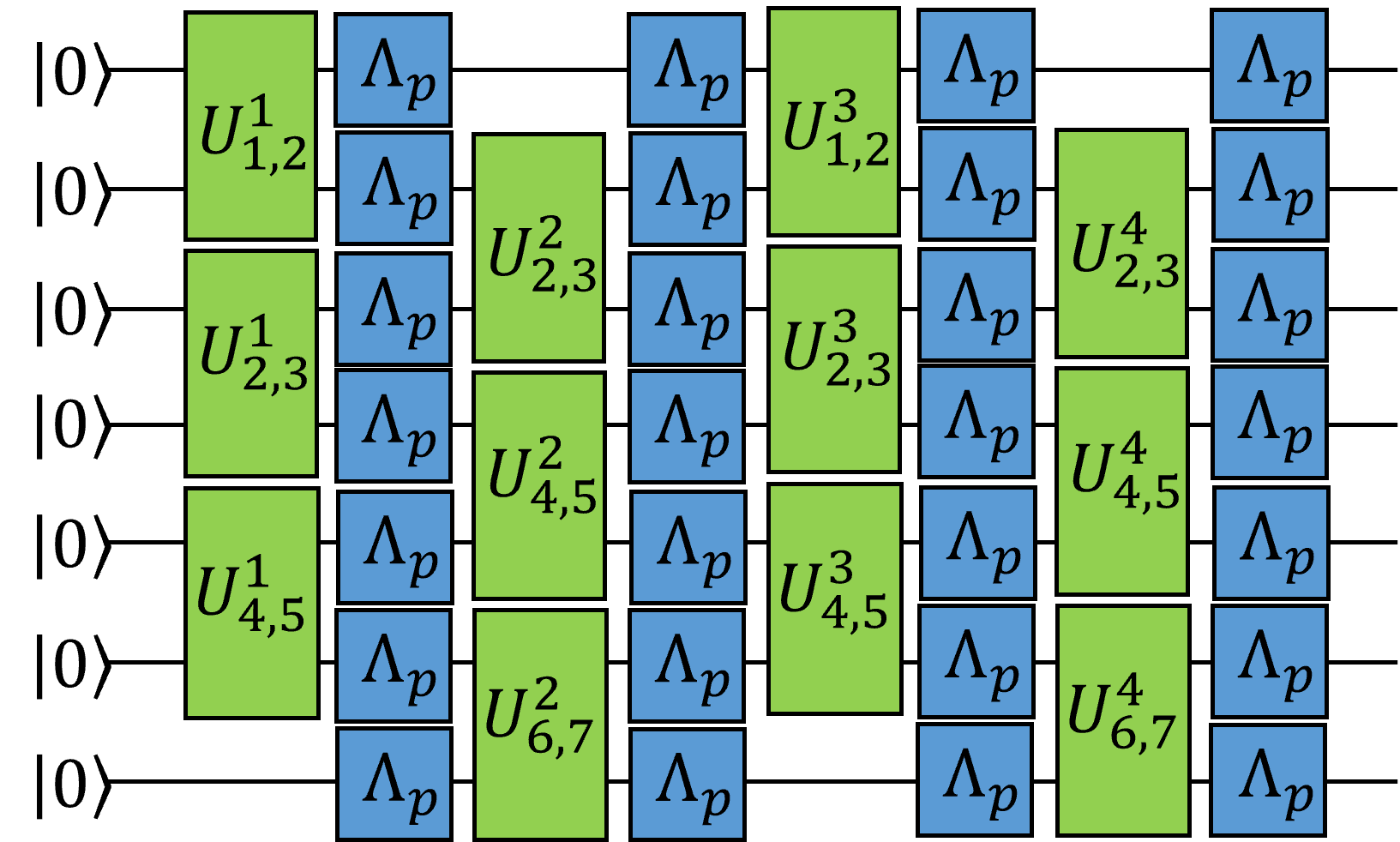}
  \caption{A diagram for LRCs with depolarizing noise. Here, each $U^{i}_{j,j+1}$ is independent random two-qubit gate, and $\Lambda_p$ is a single-qubit depolarizing noise defined by Eq.~\eqref{def_DepolarizingChannel}.}
  \label{pic_circ_LRCwithNoise}
\end{figure}

\begin{figure*}[tb!]
  \begin{minipage}[b]{0.45\textwidth}
    \centering
    \includegraphics[width=\textwidth]{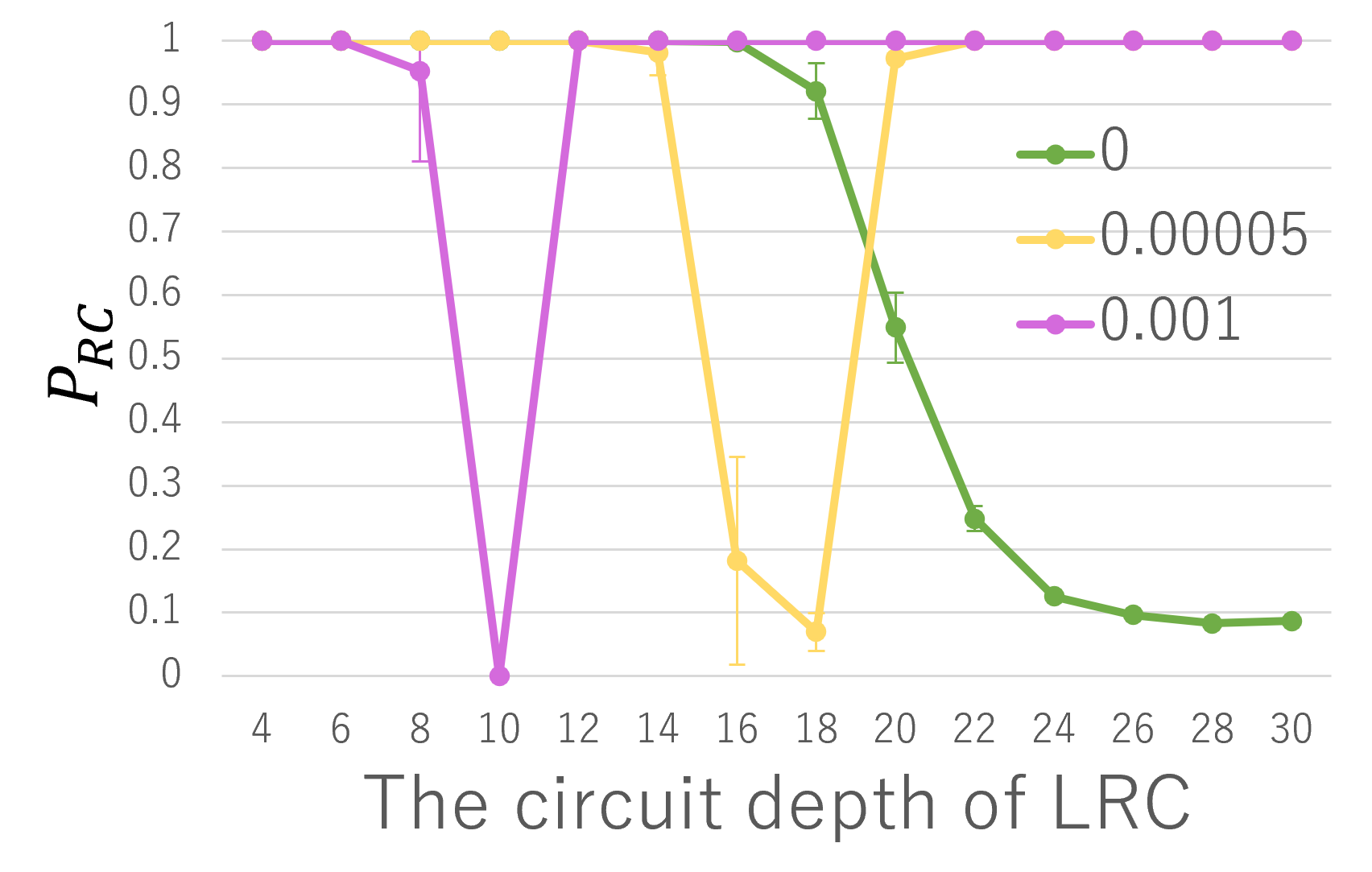}
  \end{minipage}
  \begin{minipage}[b]{0.45\textwidth}
    \centering
    \includegraphics[width=\textwidth]{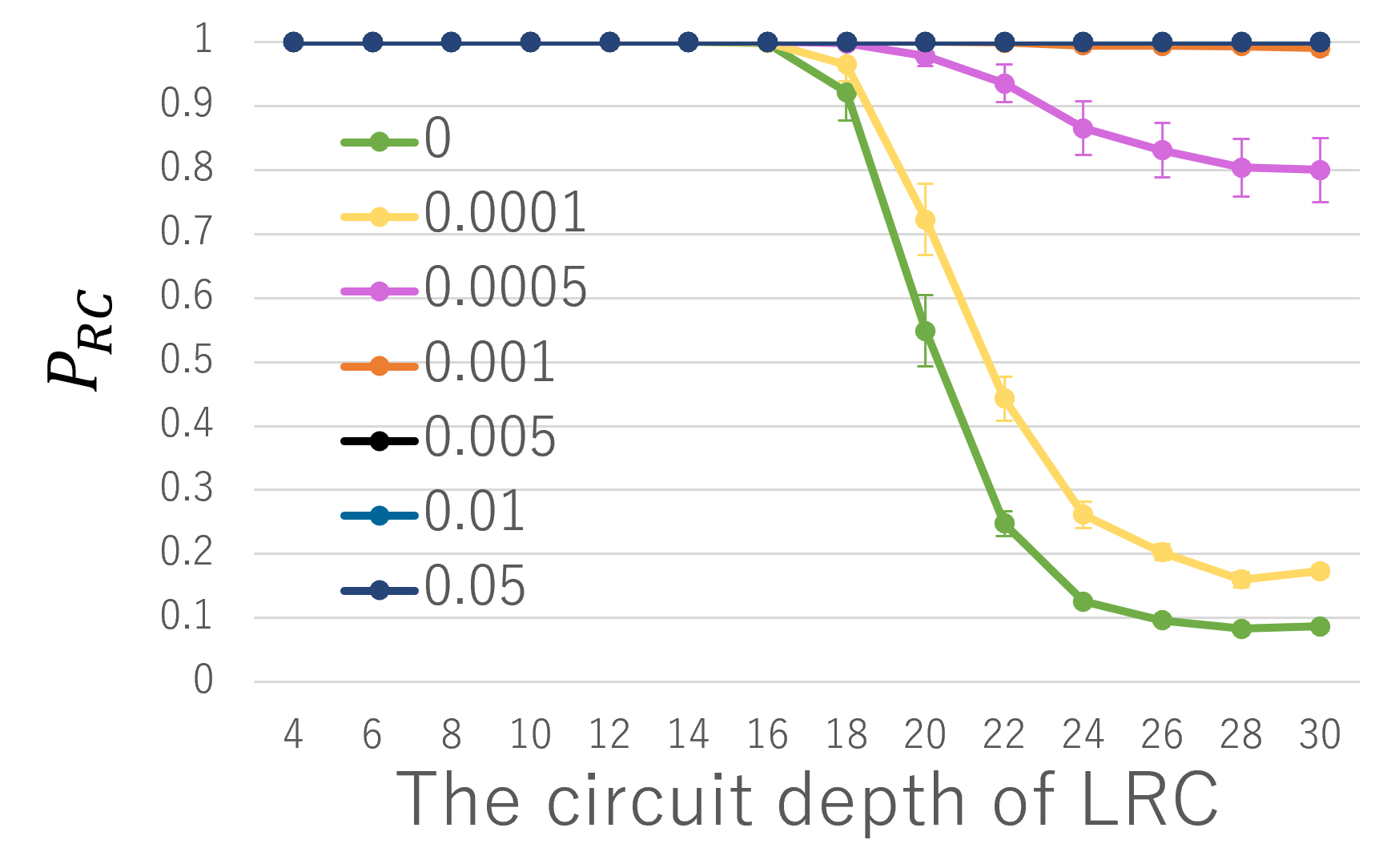}
  \end{minipage}
  \caption{Classification results of the NN classifier with $k'=4$ for the data set generated by ${\sf noisyLRC}(D, p)$ (left-hand side) and for ${\sf monitLRC}(D, p)$ (right-hand side) in $7$-qubit systems, where $P_{RC}$ is defined by Eq.(\ref{eq def P RC}). In the left figure, the graphs in green, yellow and pink correspond to $P_{RC}$ of ${\sf noisyLRC}(D, p)$ with $p=0, 0.00005, 0.001$, respectively. In the right figure, the graphs in green, yellow, pink, orange, black, blue, navy correspond to $P_{RC}$ of  ${\sf monitLRC}(D, p)$ with $p=0, 0.00001, 0.00005, 0.001, 0.005, 0.01, 0.05$, respectively.} 
  \label{pic_7q_application_s.o.r}
\end{figure*}

When LRCs are experimentally implemented, they are inevitably noisy. We here consider such noisy LRCs and see if the NN classifier detects the noise. 
As a toy model of noisy LRCs, we consider the LRCs with single-qubit depolarizing noise (see FIG.~\ref{pic_circ_LRCwithNoise}), which is commonly used in the literature~\cite{B2018, G2019}. The single-qubit depolarizing channel is defined by 
\begin{eqnarray} \label{def_DepolarizingChannel}
  \Lambda_p (\rho) &=& (1-p)\rho + p \frac{I}{2}
%  &=& (1-p)\rho + \frac{1}{3}p(X \rho X^\dagger + Y \rho Y^\dagger + Z \rho Z^\dagger) , \nonumber
\end{eqnarray}
where $I$ is the identity operator on one qubit. 
We denote by ${\sf noisyLRC}(D, p)$ a random dynamics generated by the noisy LRCs with depth $D$ and depolarizing parameter $p$. Note that this is, in general, not unitary dynamics, but our classifiers can be applied since they use only measurement probabilities.

We generate the data set of $J_{est}(h_1,\dots,h_k,k'|N_u,N_s)$ for ${\sf noisyLRC}(D, p)$ from a classical simulation. By inputting the data set to the NN classifier with $k'=4$, we obtain $P_{RC}$ as a function of the depth of the circuit, which is shown in the left-hand side of FIG.~\ref{pic_7q_application_s.o.r}.
It is clear that, even for small amount of noise such as $p \leq 10^{-3}$, $P_{RC}$ varies in the manner different from the case of noiseless LRCs. In particular, there is a `dip' as the depth increases: the probability becomes small at certain depth but increases again for larger depth. 

We can understand this `dip' as follows. 
When the depth is sufficiently shallow, ${\sf noisyLRC}(D, p)$ is not yet noisy and $P_{RC}$ behaves similarly to ${\sf LRC}(D, p)$.  
At certain depth, the noise brings significant changes to the output probability distribution of ${\sf noisyLRC}(D, p)$.
Since the noise is depolarizing, the output becomes more uniform, which is to some extent similar to the effect brought by a Haar random unitary. This is why the classifier judges ${\sf noisyLRC}(D, p)$ with a certain depth $D$ a Haar random unitary, and a `dip' appears.
As the depth further increases, the output distribution of ${\sf noisyLRC}(D, p)$ becomes more and more uniform. Since  the output probability distribution of a Haar random unitary is given by the Porter-Thomas distribution~\cite{B2018}, which differs from a completely uniform one, the classifier is able to detect the difference between ${\sf noisyLRC}(D, p)$ and a Haar random unitary in the limit of large $D$.
This intuition is confirmed by the Principal Component Analysis (PCA). See Subec.~\ref{SS:noiseLRC} for more details.

We hence conclude that the NN classifier is quite sensitive to the presence of noise in LRCs, which opens a possible application of the classifier to verify noisy quantum devices.
%In FIG.~\ref{pic_10q_5algs}, we have presented only the case with small noise. The results for more noisy LRCs are provided later in Sec.~\ref{sec of application}, where $P_{RC}$ behaves differently but succeeds to detect the presence of noise.

\subsubsection{Monitored LRCs}

\begin{figure}[tb!]
  \centering
  \includegraphics[width=0.35\textwidth]{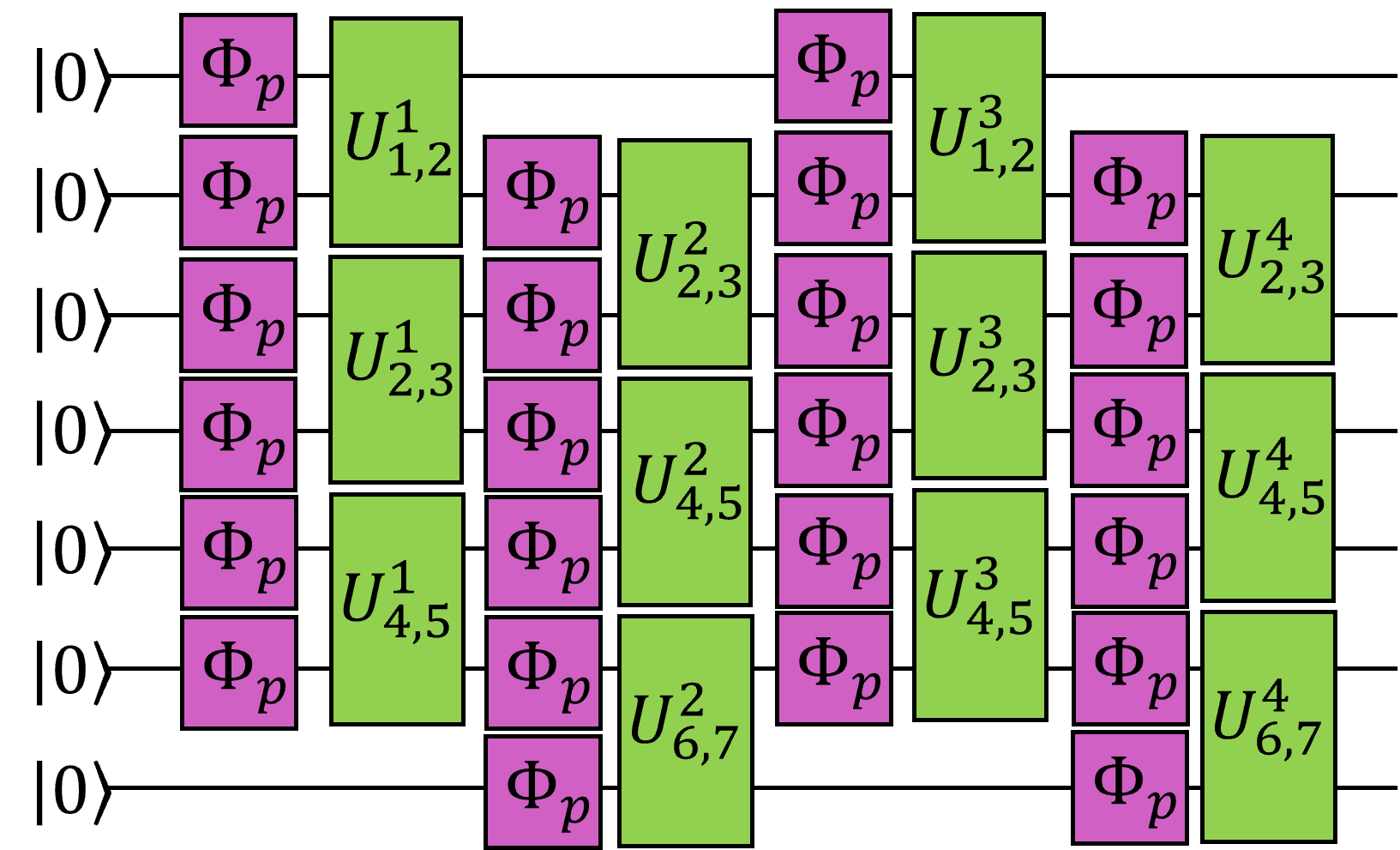}
  \caption{A diagram of monitored LRCs. Each $\Phi_p$ represents a single-qubit measurement applied with probability $p$. The measurement outcomes are all recorded, so that the output state of the circuit is pure, which is dependent on the whole measurement outcomes. Each $U^{i}_{j,j+1}$ is independent random two-qubit gate. }
  \label{pic_circ_LRCwithMeasurement}
\end{figure}

We next apply the NN classifier to the monitored LRCs. The monitored LRCs are defined as LRCs, where each qubit in each layer is measured in the computational basis with a fixed probability $p$ (see FIG.~\ref{pic_circ_LRCwithMeasurement}).  To avoid the confusion, we refer to $p$ as the measurement ratio.
An important assumption is that we know the measurement outcome. Hence, at the end of the circuit, we obtain a pure state dependent on the whole measurement outcomes. We also assume that the measurement is not applied in the last step. We denote this monitored LRCs with depth $D$ and the measurement ratio $p$ by ${\sf monitLRC}(D,p)$. 

The monitored LRCs were proposed in~\cite{LCF2018} and were shown to have a novel phase transition of dynamics depending on $p$. The transition is named a \emph{measurement-induced phase transition} (MIPT) and can be witnessed by the average entanglement entropy of the system: for small $p$, entanglement grows quickly to all the system, while it remains constant for large $p$. 
From the fact that LRCs with sufficient depth are unitary designs, studying MIPTs in terms of designs will help deeply understanding MIPTs from the perspective of randomness. However,
the entanglement entropy and other functions exhibit MIPTs \cite{PhysRevB.101.104301,PhysRevX.10.041020,PhysRevB.101.104302} have no direct relation with a $t$-design. Further, since a $t$-design is  about unitary dynamics and cannot be directly applied to non-unitary dynamics, it is not straightforward to investigate the monitored LRCs in terms of $t$-designs. Our classifier is constructed based on a necessary condition of $t$-designs and is applicable to non-unitary random dynamics. 
Hence, by applying our method to the monitored LRCs, we  are able to obtain insights into the relation between MIPT and the randomness in terms of $t$-designs.

By classically simulating ${\sf monitLRC}(D,p)$, and by inputting the measurement outcomes into the NN classifier, we obtain $P_{RC}$ as depicted in the right-hand side of FIG.~\ref{pic_7q_application_s.o.r}.
We observe that the probability $P_{RC}$ increases as the measurement ratio $p$ increases, and even for a relatively small $p$ such as $p = 10^{-3}$, the output of ${\sf monitLRC}(D,p)$ is classified as RCs with unit probability.
This can be intuitively explained because each measurement in ${\sf monitLRC}(D,p)$, recalling that it is in the computational basis, locally resets the depth of the circuit. This may lead an effective depth shallower, and the classifier outputs ${\sf RC}$ even for large $D$. 
It is, hence, reasonable to conclude that it is unlikely that MIPTs can be observed from the classification result. Since the classification is about if the dynamics forms a $t$-design, this may suggest that MIPTs are possibly not a transition of randomness in terms of $t$-designs.

Although the NN classifier fails to witness MIPTs, it is of practical use in the realization of experiments of  MIPTs. In practical implementations of the monitored LRCs, the effect of noise is inevitable, which rapidly increases the entropy of the whole system. This makes it difficult to detect MIPTs since the entropy induced by the noise, rather than entanglement entropy, will quickly become dominant in the system. Hence, toward the realizations of MIPTs, it is important to keep the system as noiseless as possible. For this purpose, the NN classifier can contribute: as we have observed in the previous section, the NN classifier is very sensitive to the presence of noise. Hence, by checking the output data of the monitored LRCs using the classifier, it is possible to check if the whole system remains noiseless.

\section{Organization of the rest of the paper \label{Sec:OrR}}

In the remaining of the paper, we explain our analysis in detail. We provide the method to generating feature vectors in Sec.~\ref{S:FVdetail}, which are the input of the supervised learning as well as the classifiers trained by the learning. 
The method of supervised learning and evaluations are provided in Sec.~\ref{creation and evaluation about the discriminator}, which are based on random Clifford ${\sf RC}$ and the Haar measure ${\sf Haar}$. We then apply the constructed classifiers to the random dynamics that are substantially different from ${\sf RC}$ and ${\sf Haar}$. The detailed analysis on local random circuits is given in Sec.~\ref{S:CLRCsRDCs}, and non-unitary random dynamics in Sec.~\ref{sec of application}.

\section{Generating feature vectors} \label{S:FVdetail}

We here explain how we generate the data set for a given ensemble ${\sf U}$ of unitaries on $N_q$ qubits.
We sample $N_u$ unitaries from ${\sf U}$. From each sampled unitary $U_i$, we generate a state 
\begin{equation}
  \ket{\psi_{out}^{(i)}} = U_i  P \ket{0}^{\otimes N_q},
\end{equation}
where $P$ is pre-processing and is applied if necessary.
More concretely, we use non-trivial $P$ only in the training stage for making the training more effective, which will be later explained.

Each state $\ket{\psi_{out}^{(i)}}$ is measured in the computational basis, resulting in a single measurement outcome.
For each $U_i$, we repeat preparing and measuring the state $N_s$ times, from which we estimate $\bra{\psi_{out}^{(i)}} Z_{h_1} \otimes \dots \otimes Z_{h_k} \ket{\psi_{out}^{(i)}}$. Denoting the measurement outcome on the $h_p$th qubit in the $l$th trial for $U_i$ by $x(i, l, p) \in \{ 0, 1 \}$, the estimated value is given by
\begin{equation}
\bra{\psi_{out}^{(i)}} Z_{h_1} \otimes \dots \otimes Z_{h_k} \ket{\psi_{out}^{(i)}}_{est}
:=
\frac{1}{N_s}\sum^{N_s}_{l=1} \prod_{p=1}^{k} (-1)^{1-x(i, l, p)} \label{Eq:111}
\end{equation}
From this, we obtain an estimator of the $k'$th moment of $k$-bit correlations $J_{est}(h_1,\cdots, h_k,k'| N_u,N_s)$:
\begin{multline} \label{eq_bitcorr_2 est2}
J_{est}(h_1,\cdots, h_k,k'| N_u,N_s)\\:=  \frac{1}{N_u} \sum^{N_u}_{i=1} \bigl( \bra{\psi_{out}^{(i)}} Z_{h_1} \otimes \dots \otimes Z_{h_k}  \ket{\psi_{out}^{(i)}}_{est} \bigr)^{k'}
\end{multline}
These estimated values are the input of supervised learning as well as classifiers.

Note that 
\begin{align}
& J_{est}(h_1,\cdots, h_k,k'| N_u,N_s) \nonumber \\
\rightarrow & \mathbb{E}_{U_i \sim {\sf U}} \bra{\psi_{out}^{(i)}} Z_{h_1} \otimes \dots \otimes Z_{h_k} \ket{\psi_{out}^{(i)}}^{k'} \nonumber\\
=& {\rm Tr} \left[Z_{h_1} \otimes \dots \otimes Z_{h_k}\cdot G^{(k')}[{\sf U}]\left( \left( P\left(\ket{0}\bra{0}\right)^{\otimes N_q}P^\dagger \right)\right)
\right]
\end{align}
in the limit of $N_s, N_u \rightarrow \infty$. Due to Eq.~\eqref{def of t-design}, if ${\sf U}$ is a $k'$-design, the last line is equal to the average ober ${\sf Haar}$.
This fact suggests that the classifier constructed based on 
$J_{est}(h_1,\cdots, h_k,k'| N_u,N_s)$ is useful for classifying $k'-1$ design and $t$-design with $t\ge k'$.
In most of our numerical studies, $k'$ is fixed to, e.g., $4$, but it can be more than one, such as $k'\in \{4,6,8,10 \}$.

In the context of machine learning, an estimated value $J_{est}(h_1,\cdots, h_k,k'|N_u,N_s)$ is called a \emph{feature}, and a set of estimated valued over $1\le \forall k\le N_q$ and $1\le \forall h_1 < \cdots < \forall h_k \le N_q$ is called a \emph{feature vector}. A \emph{data set} is a collection of feature vectors over $N_u$ and $N_s$, which is generated by repeating the above procedure many times.
When we generate a data set, we apply $z$-score normalization. That is, for each feature in a data set, we compute the average and the standard deviation over the set. By using this average and standard deviation, we normalize the set such that the average over all features of the new data set is $0$, and the standard deviation is $1$.

\section{Training and evaluations of classifiers}\label{creation and evaluation about the discriminator}

In this Section, we provide details of the learning methods based on the feature vectors genarated in the way in Sec.~\ref{S:FVdetail}. We explain how to generate data sets in Subsec.~\ref{SS:GDS} and training and evaluating methods in Subsec.~\ref{SS:TM}. 
We also check the performance of classifiers based on different moments of the data set in Subsec.~\ref{SS:Moments}.

\subsection{Generating data sets} \label{SS:GDS}

In the training stage, we use two random dynamics, ${\sf RC}$ and ${\sf Haar}$. 
While we set $P=I$ for ${\sf Haar}$, we set $P = V$ for ${\sf RC}$, where $V$ is a unitary chosen from ${\sf Haar}$ Haar random unitary and is fixed for all feature vectors. This is to randomize the basis and is necessary for learning the structures of a $3$-design, not those specific to RCs. See Appendix.~\ref{Fixed_basis} for the results derived without the pre-processing.

For each of the random dynamics ${\sf RC}$ and ${\sf Haar}$, we classically simulate the measurement outcomes of $\ket{\psi_{out}^{(i)}}$ in the computational basis, which is performed using the library qulacs~\cite{suzuki2020qulacs} and the normal form of Clifford operators~\cite{lmcs:1570}, and obtain a set of feature vectors.
We generate three types of data sets, training, validation, and test data sets consisting of $O_{train}$, $O_{valid}$, $O_{test}$ feature vectors, respectively. 
For each data set, one half is generated from ${\sf RC}$ and the other half from ${\sf Haar}$. The $z$-score normalization is applied to all the data set.

The number of qubits we use in the analysis is fixed to either $7$ or $10$ qubits. The former is mainly used to investigate the learning in detail.
The parameters of the training data set for $7$-qubit and $10$-qubit systems are summarized in TABLE \ref{table_paras_teacherdata}. 
These numbers are chosen as large as possible with our computational resource (see Appendix.~\ref{append_envs}).

\subsection{Training and evaluating methods} \label{SS:TM}

\begin{table}[tb!] 
  \centering
  \caption{Parameters of training data}
  \begin{tabular}{ccc}
    \hline
    Parameter & $7$-qubit & $10$-qubit\\
    \hline \hline
    $N_q$ & 7 & 10\\
    $O_{train}$ & 20000 & 20000 \\
$O_{valid}$ & 20000 & 20000\\
    %\# of test data $O_{test}$ & 10000 \\
    $N_u$ & 10000 & 20000\\
    $N_s$ & 10000 & 10000\\
    Range of $k$ & $\{1,\cdots,7\}$  & $\{1,\cdots,10\}$\\
    Range of $k'$ & $\{4\}$ & $\{4\}$ \\
    \hline
    \label{table_paras_teacherdata}
  \end{tabular}
\end{table}

We use five learning algorithms: linear regression (LR), linear support vector machine (LSVM), support vector machine (SVM), random forest (RF), and neural network (NN). We implement NN using Keras~\cite{chollet2015keras} with TensorFlow~\cite{tensorflow2015-whitepaper} as the back end and the others using scikit-learn~\cite{scikit-learn}. A brief description of these algorithms and detail settings, e.g. hyper parameters, are provided in Appendix.~\ref{ML_algos}.

In the training and validation steps, we construct ten classifies for each learning algorithm .
To construct each classifier, we unite the training and validation data set and randomly divide it into two sets of the same size. 
The number of vectors generated by ${\sf RC}$ and that by ${\sf Haar}$ can be different in the divided sets. One set is used for training, and the other is for validation. To each data set, we apply the z-score normalization.
We then use the test data set to check the performance of each classifier constructed by each algorithm, which is determined by the average over the $10$ classifiers.

\setlength{\tabcolsep}{6pt} % Default value: 6pt
\renewcommand{\arraystretch}{1.2} % Default value: 
\begin{table}[tb!]
  \centering
  \caption{Average success probabilities and standard deviations of classification by various Learning Algorithms. Here, the number $N_q$ of qubits is set to $7$.}
  \begin{tabular}{c c c c}
    \hline
    Algorithms & training & validation & test \\
    \hline  \hline
    LR   & 0.535$\pm$0.004 & 0.504$\pm$0.002 & 0.502$\pm$0.004 \\
    LSVM & 0.535$\pm$0.003 & 0.505$\pm$0.003 & 0.502$\pm$0.003 \\
    SVM  & 0.886$\pm$0.001 & 0.845$\pm$0.002 & 0.847$\pm$0.001 \\
    RF   & 1.000$\pm$0.000 & 0.803$\pm$0.002 & 0.806$\pm$0.002 \\
    NN   & 0.856$\pm$0.004 & 0.845$\pm$0.002 & 0.847$\pm$0.001 \\
    \hline
    \label{table_asp_7q}
  \end{tabular}
\end{table}

We present the average success probabilities and its standard deviations of the classification of the training, validation, and test data sets for each learning algorithm on a $7$-qubit system in TABLE~\ref{table_asp_7q} (see TABLE ~\ref{table_10q_ASP} in the case of $10$ qubits).
From TABLE \ref{table_asp_7q}, we observe that the average success probabilities for the classification by SVM, RF an NN for $7$ qubits are more than 80\%, which are much higher than those for $10$ qubits. 
This fact suggests that the size of the data set necessary for achieving high accuracy increases as the dimension of the system is large.
On the other hand, we observe that learning by LR and LSVM fails and the success probability is approximately 50\%. 

By comparing the results for $7$ with $10$ qubits, one may notice that all algorithms are consistent except the classifier constructed by RF, which succeeds learning for $7$ qubits but fails for $10$ qubits. This is understood as follows. RF is a classifier that achieves strongly non-linear classification with ensemble learning and perfectly classify the training data for $7$ and $10$ qubits. On the other hand, this tends to over-fits the training dataset and cannot classify the validation and test datasets as the complexity of datasets increases. While the performance of RF may be improved by hyper-parameter tuning in this case, such tuning are beyond the scope of the paper and we left it as future work.

\subsection{Learning from various moments} \label{SS:Moments}

\begin{figure}[tb!]
  \centering
  \includegraphics[width=8cm]{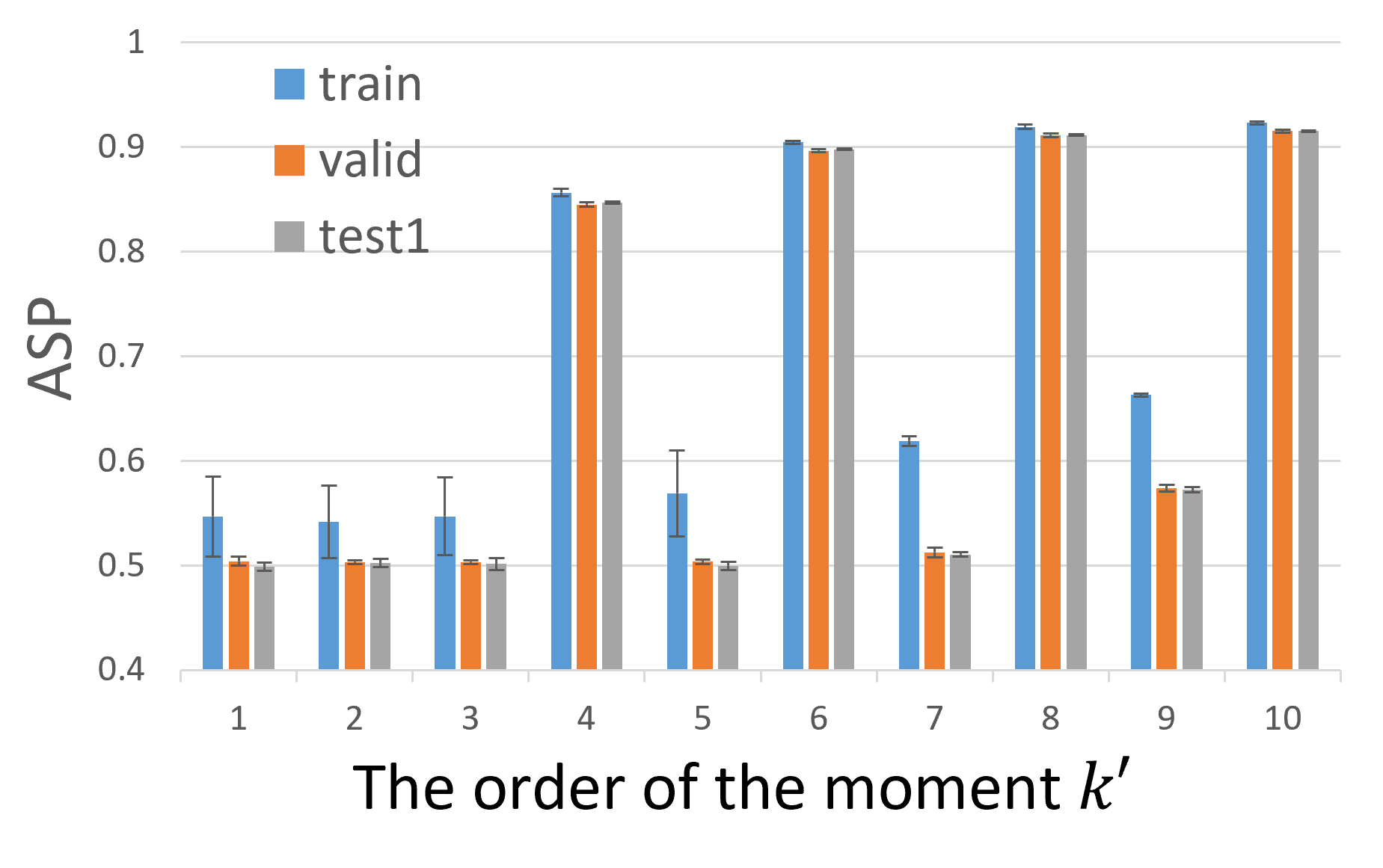}
  \caption{Classification results of the NN classifiers with various $k'$. The vertical axis is the average success probability (ASP) of classification of the training, validation, and test data. A bar graph and an error bar represent an average and a standard deviation over $10$ classifiers, respectively.}
  \label{pic_test1and2_1to10}
\end{figure}

We next show that the performance of the classifiers is improved when the moment $k'$ is set to a higher value in the learning data set of $J_{est}(h_1,\cdots, h_k,k'| N_u,N_s)$. See Eq.~\eqref{eq_bitcorr_2 est2} as well. We here use the NN algorithm since it is more reliable and faster than the others. The result of the average success probabilities for $7$ qubits is summarized in FIG.  \ref{pic_test1and2_1to10}.

We first observe that the average success probability for the validation and test data sets is about $50$\% for $k'\le 3$, which clearly implies that the failure of learning. This is simply because both ${\sf RC}$ and ${\sf Haar}$ are $3$-designs. Hence, as far as concerning the property of degree at most $3$, it is not possible to distinguish them by any means.

One may be surprised with the fact that learning from the data set with odd $k'$ does not work well even for $k' \geq 4$. Although the accuracy of odd $k'$ improves as $k'$ gets larger, it is much smaller than the values compared to the case of even $k'$.
The reason for this behavior arises from the fact that $J_{est}(h_1,\cdots, h_k,k'| N_u,N_s)$ is the average over positive and negative quantities when $k'$ is odd (see Eq.~\eqref{eq_bitcorr_2 est2}). This makes the value of $J_{est}(h_1,\cdots, h_k, k'| N_u, N_s)$ possibly almost zero.
Taking into account the z-score normalization, a statistical error in training data may be significantly large. 
Thus, more precision in the numerical simulations is required. For this reason, the results are very sensitive to the precision of numerics and become unstable.

%\red{unstable if $k'$ is large?}

\section{Classifications of random unitary dynamics} \label{S:CLRCsRDCs}

\begin{figure*}[tb!]
\centering
  \begin{minipage}[tb]{0.43\textwidth}
  \includegraphics[width=\textwidth]{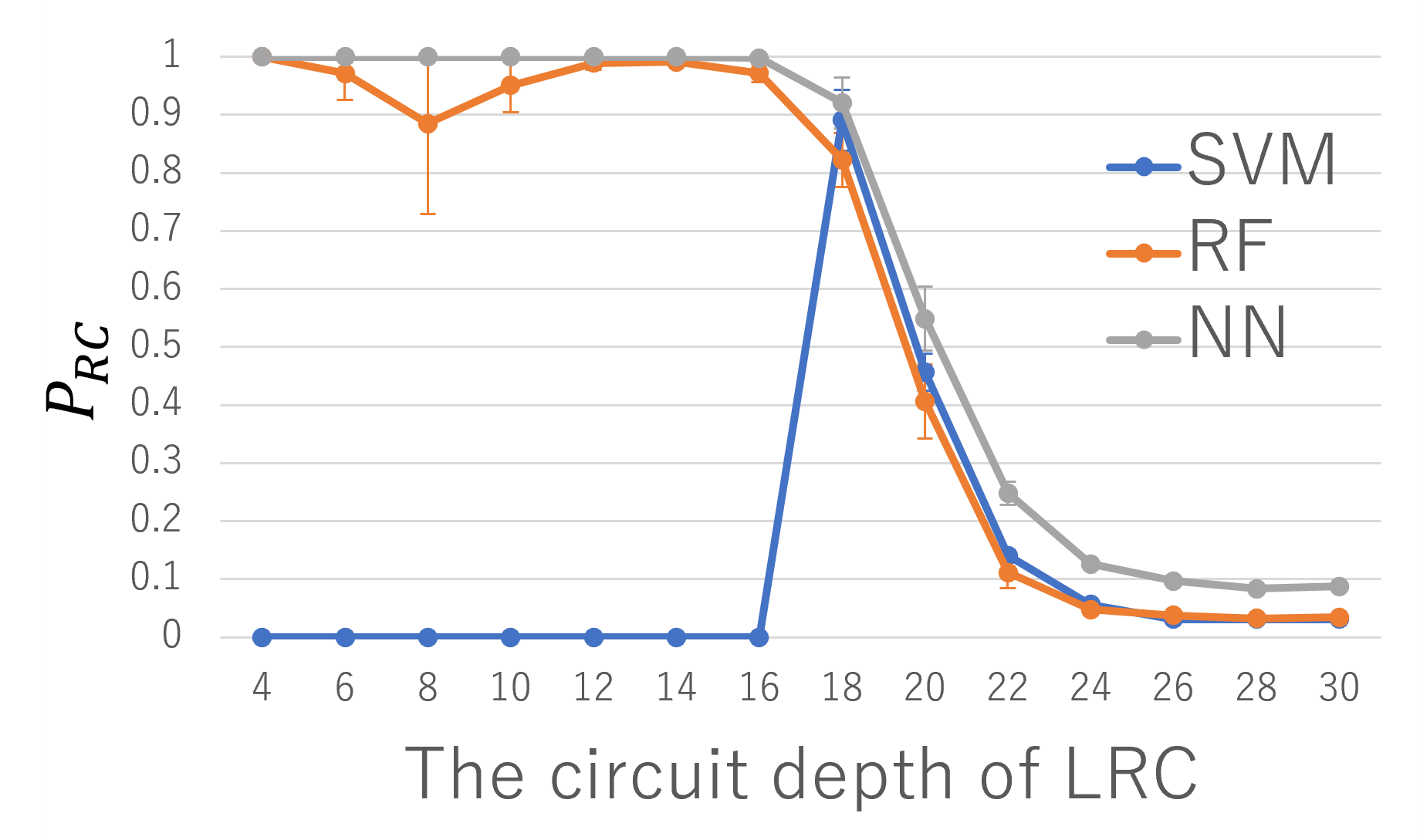}
  \end{minipage}
   \begin{minipage}[tb]{0.45\textwidth}
	\includegraphics[width=\textwidth]{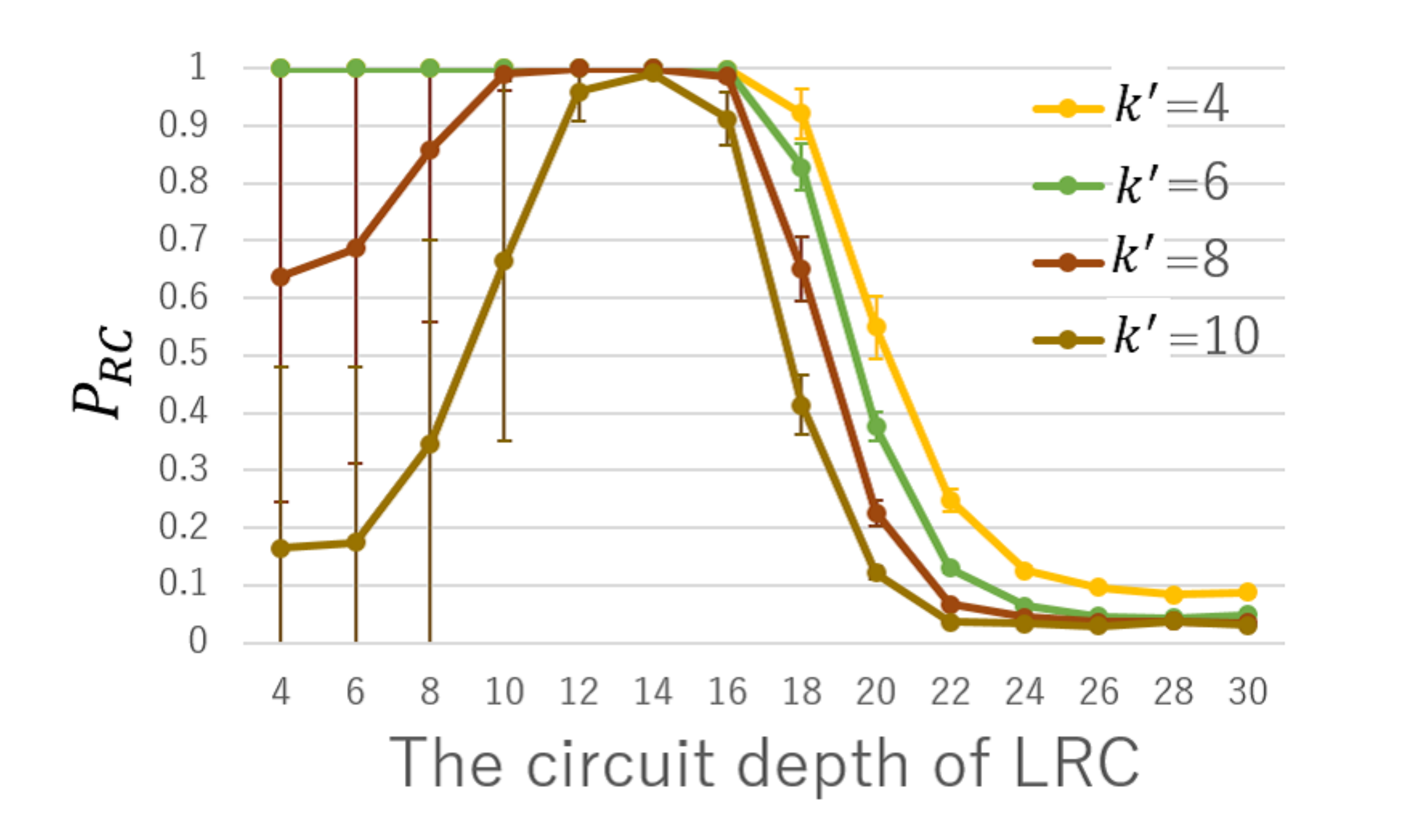}
  \end{minipage}
    \caption{(Left) Classification results of the test datasets generated by ${\sf LRC}(D)$ of $7$ qubits. The vertical axis represents the probability $P_{RC}$ that the classifier output ${\sf RC}$, which defined by Eq.(\ref{eq def P RC}), and the horizontal axis the circuit depth of LRCs. (Right) Classification results of the test datasets generated by ${\sf LRC}(D)$ for four classifiers constructed by the data set with  different moments $k'=4,6,8,10$. In both figures, the dot is the average and the error bar is the standard deviation of the results obtained from $10$ classifiers.
     }
    \label{pic_LRC_1to7}
\end{figure*}

As an application of the classifiers obtained in the previous section, we investigate a classification problem of random circuits by the classifiers. We first consider ${\sf LRC}(D)$, and then ${\sf RDC}(I)$.
Since both show transitions in terms of $t$-designs as the depth or the number of iterations increase (see Subsec.~\ref{subsec Random quantum circuits}), they can be used to check if the classifiers succeed to detect the growth of complexity in the sense of designs. 

In the following, we focus on $7$-qubit systems and use the classifiers based on SVM, RF, and NN. They are the classifiers that succeed to learn the features of ${\sf RC}$ and ${\sf Haar}$ for $7$ qubits.

\subsubsection{Methods}

To test how ${\sf LRC}(D)$ and ${\sf RDC}(I)$ are characterized by the classifiers, we generate the data set, consisting of $1000$ feature vectors, by classically simulating the circuits.
As mentioned before, we set the pre-processing $P$ as $I$.
The moment $k'$ is fixed to $4$ unless specified.
For each data set, we evaluate the probability $P_{RC}$ for the classifier to output ${\sf RC}$. To avoid rare cases of learning, we compute the probability for $10$ classifiers of the same kind and take the average.

Note that we have used ${\sf RC}$ for constructing classifiers, which is an exact $3$-design but not a higher-design. Hence, even though we label the probability as $P_{RC}$, it is nothing to do with ${\sf RC}$ and is rather expected that it is the probability about whether a given random circuit is at most a $3$-design or a higher-design.

\subsubsection{Classifying LRCs}

For classifying ${\sf LRC}(D)$, we generate data sets for $D = 4, \dots, 30$, from which we obtain $P_{RC}$ as a function of the depth $D$. The result is shown in the left-hand side of FIG.~\ref{pic_LRC_1to7} (see FIG.~ \ref{pic_10q_5algs} for $10$ qubits).
We have plotted the results for the circuits with even depth since ${\sf LRC}(D)$ with even and odd depths $D$ have different structures due to the periodicity of the layers.

We observe that all of the SVM, RF, and NN classifiers show characteristic behaviors around the depth between $18$ and $20$, which indicates that all of them are able to detect the growth of circuit complexity, in the sense of $3$-designs, in ${\sf LRC}(D)$.

Despite the fact that all classifiers seem to capture the growth of complexity, the value of $P_{RC}$ of the SVM classifier is different from the other two in the regime of shallow depth. This should be for the reason discussed in the second last paragraph in Subsec.~\ref{SS:LRCSum}. That is, since the LRCs with shallow depth are not $3$-designs and far from both RCs and a Haar random unitary, the output of classifiers cannot be reliable. 
This instability of the SVN classifier is also observed by comparing the results for $7$ and $10$ qubits (the left-hand side of FIG.~\ref{pic_LRC_1to7} and FIG.~ \ref{pic_10q_5algs}), where the SVM classifiers determines the shallow LRCs as a Haar random unitary for $7$ qubits, but as RCs for $10$ qubits.
It is not clear why the SVM classifier is particularly unstable, in comparison with the other two, but this fact leads us to use the NN classifiers in the main analysis of $10$ qubits as we have shown in Sec.~\ref{s.o.r}.

We finally apply the NN classifiers constructed from the data set with larger $k'$ to the data set generated by LRCs for $7$ qubits, which is shown in the right-hand side of FIG.~\ref{pic_LRC_1to7}. Similarly to the case of $10$ qubits shown in FIG~\ref{pic_kp4_6_8_10_nn} in Subsec.~\ref{SS:LRCSum}, a larger $k'$ seems to have better performance. For the reason same as the case of $10$ qubits, we expect that the depth of LRCs, at which it changes from a $3$-design to a higher-design, is between $14$ and $18$. The former is decided from the fact that the standard deviation of $P_{RC}$ becomes reasonably small at that depth, and the latter is the depth we have observed from the data set of $k'=10$. See Subsec.~\ref{SS:LRCSum} as well.

\subsubsection{Classifying RDCs}

\begin{figure*}[tbh!]
\centering
  \begin{minipage}[tb]{0.44\textwidth}
  \includegraphics[width=\textwidth]{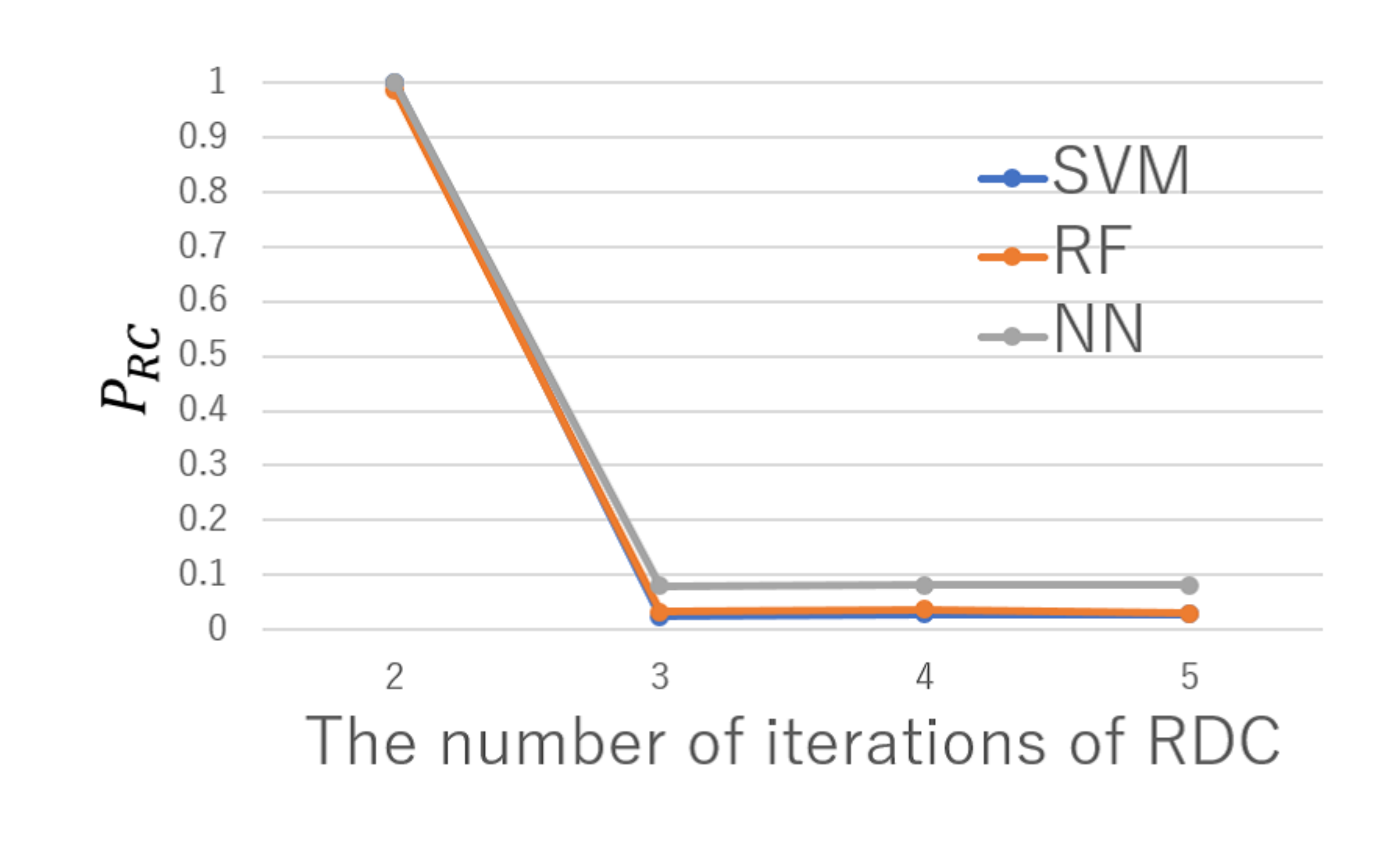}
  \end{minipage}
   \begin{minipage}[tb]{0.43\textwidth}
	\includegraphics[width=\textwidth]{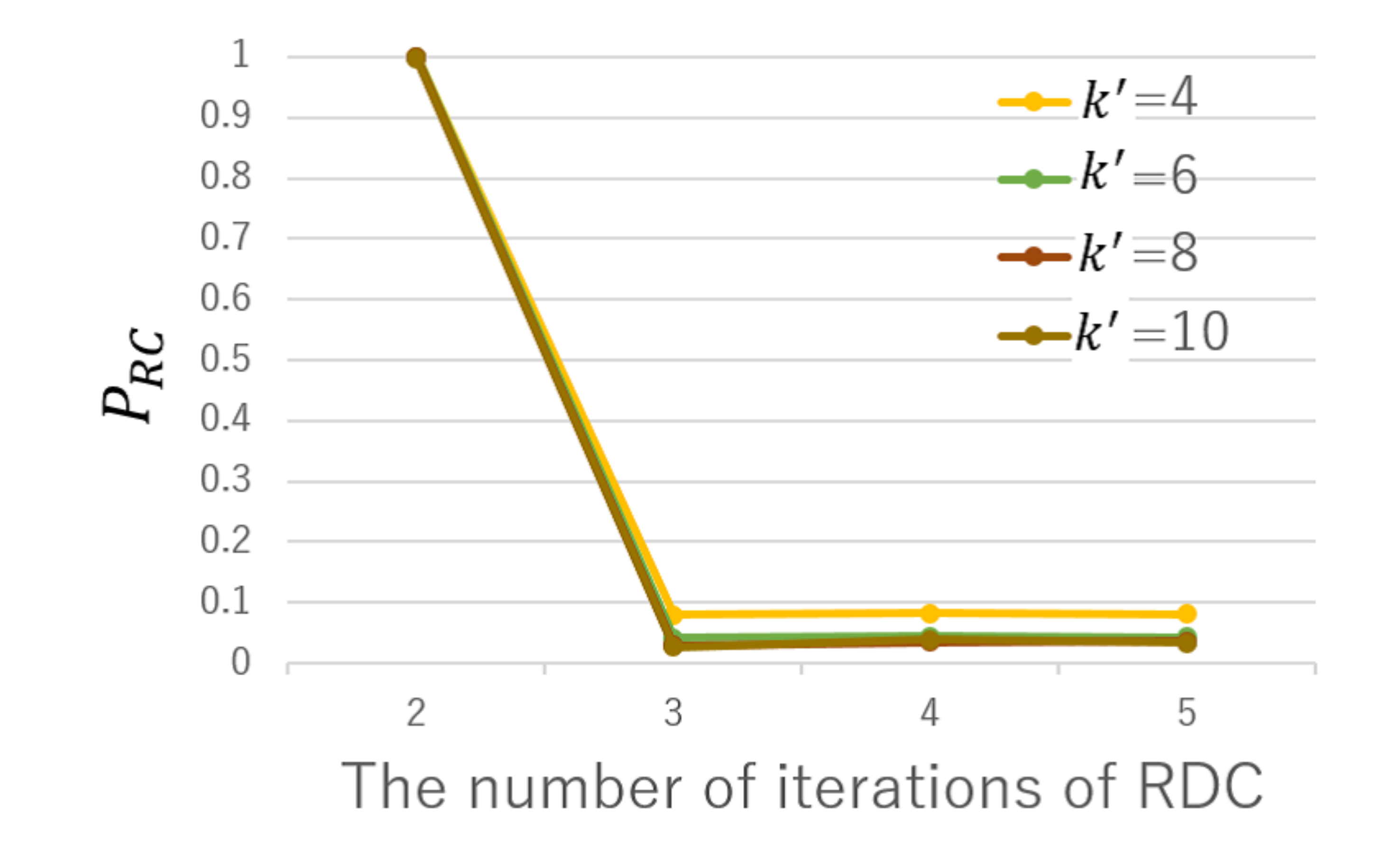}
  \end{minipage}
    \caption{
    (Left) Classification results of the test datasets generated by ${\sf RDC}(I)$ of $7$ qubits. The vertical axis represents the probability $P_{RC}$ that the classifier output ${\sf RC}$, which defined by Eq.(\ref{eq def P RC}), and the horizontal axis the number $I$ of iterations of ${\sf RDC}(I)$. (Right) Classification results of the test datasets generated by ${\sf RDC}(I)$ for four classifiers constructed by the data set with  different moments $k'=4,6,8,10$. In both figures, the dot is the average and the error bar is the standard deviation of the results obtained from $10$ classifiers. }
    \label{pic_RDC_1to7} 
\end{figure*}

We also apply our classifiers with $k'=4$ to the data set generated by ${\sf RDC}(I)$ for all $2\le \forall I \le 5$. The number of qubits is fixed to $7$.
The result is shown in the left-hand side of FIG.~\ref{pic_RDC_1to7}.
Classification results of ${\sf RDC}(I)$ by the classifiers constructed from the data set of $k' \in \{4,6,8,10\}$ are given in the right-hand side.

It is clear that ${\sf RDC}(I)$ are classified to ${\sf Haar}$ when $I \geq 3$. A known bound for $I$ such that ${\sf RDC}(I)$ forms a $t$-design is given by Eq.~\eqref{RDC depth and design}. It predicts that $I \geq 3.8$ is sufficient for ${\sf RDC}(I)$ to be a $3$-design, where a contribution from $\epsilon$ is ignored.
Thus, all of the SVM, RF, and NN classifiers constructed from any $k' \in \{4,6,8,10\}$ are consistent with the analytical bound. This supports the fact that the changing point of the dynamics from $3$- to higher-designs can be detected by the classifies, even though the dynamics is different from the dynamics used for training.

\section{Classification of non-unitary random dynamics} \label{sec of application}

We apply the NN classifier constructed from the data set generated by ${\sf RC}$ and ${\sf Haar}$ to various types of non-unitary random dynamics. We first apply it to the LRCs with depolarizing noise in Subsec.~\ref{SS:noiseLRC} and then to the monitored LRCs in Subsec.~\ref{SS:moni}. In both cases, we focus on $7$ qubits and fix the moment $k'$ to $4$.

These dynamics are not unitary, so that the concept of unitary designs cannot be directly applied. However, our classifier depends only on the measurement outcomes after the dynamics and is applicable to any dynamics, which we think is one of the advantage of our approach.

\subsection{Characterization of noisy LRCs} \label{SS:noiseLRC}
\begin{figure*}[tbh!]
\centering
  \begin{minipage}[tb]{0.35\textwidth}
    \includegraphics[width=\textwidth]{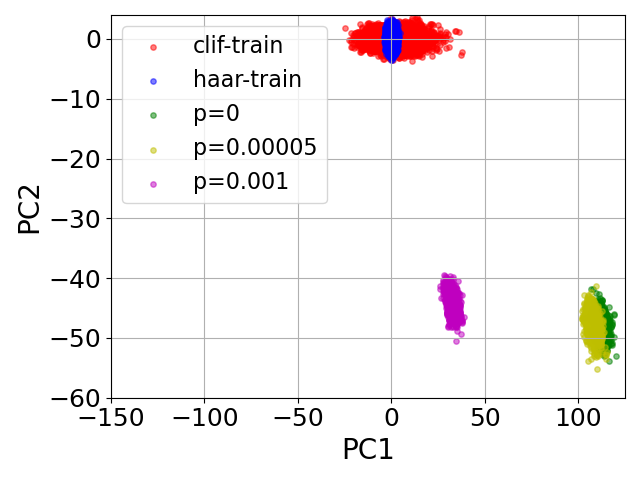}
  \end{minipage}
   \begin{minipage}[tb]{0.35\textwidth}
    \includegraphics[width=\textwidth]{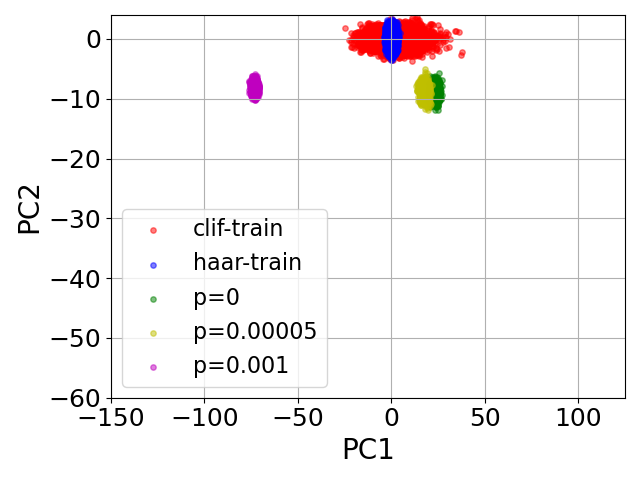}
  \end{minipage}\\
    \begin{minipage}[tb]{0.35\textwidth}
    \includegraphics[width=\textwidth]{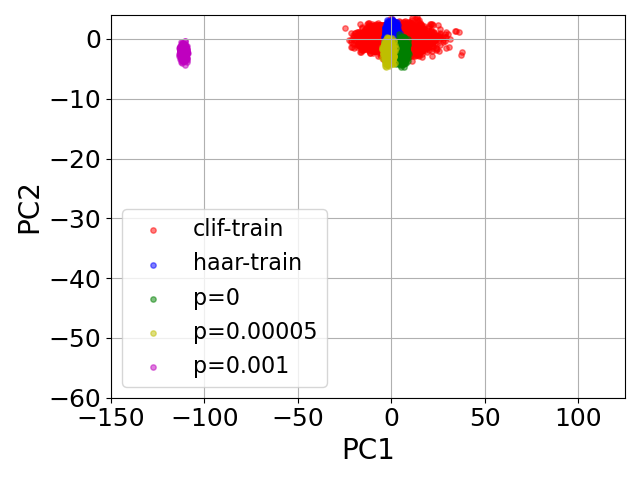}
  \end{minipage}
   \begin{minipage}[tb]{0.35\textwidth}
    \includegraphics[width=\textwidth]{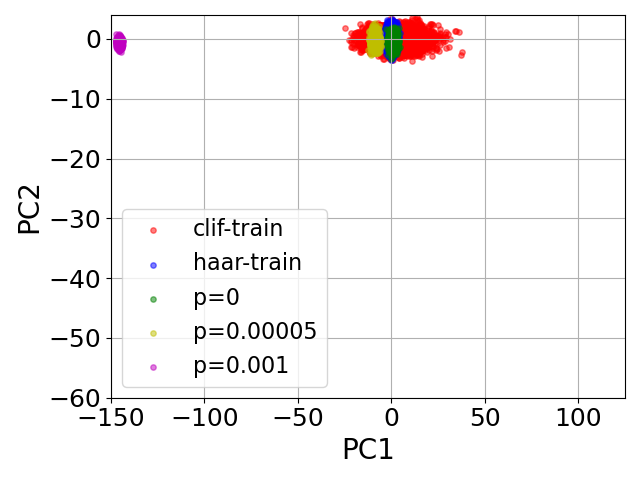}
  \end{minipage}
    \caption{The result of PCA for the data set generated by ${\sf noisyLRCs}(D,p)$ of $7$ qubits with $D= 10$ (left-top), 14 (right-top), 18 (left-bottom), and 24 (right-bottom), and $p = 0, 5 \times 10^{-5}$ and $10^{-3}$.}
  \label{pic_pca_noise_d}
\end{figure*}

\begin{figure*}[tb!]
  \begin{minipage}[b]{0.4 \textwidth}
    \centering
    \includegraphics[width=\textwidth]{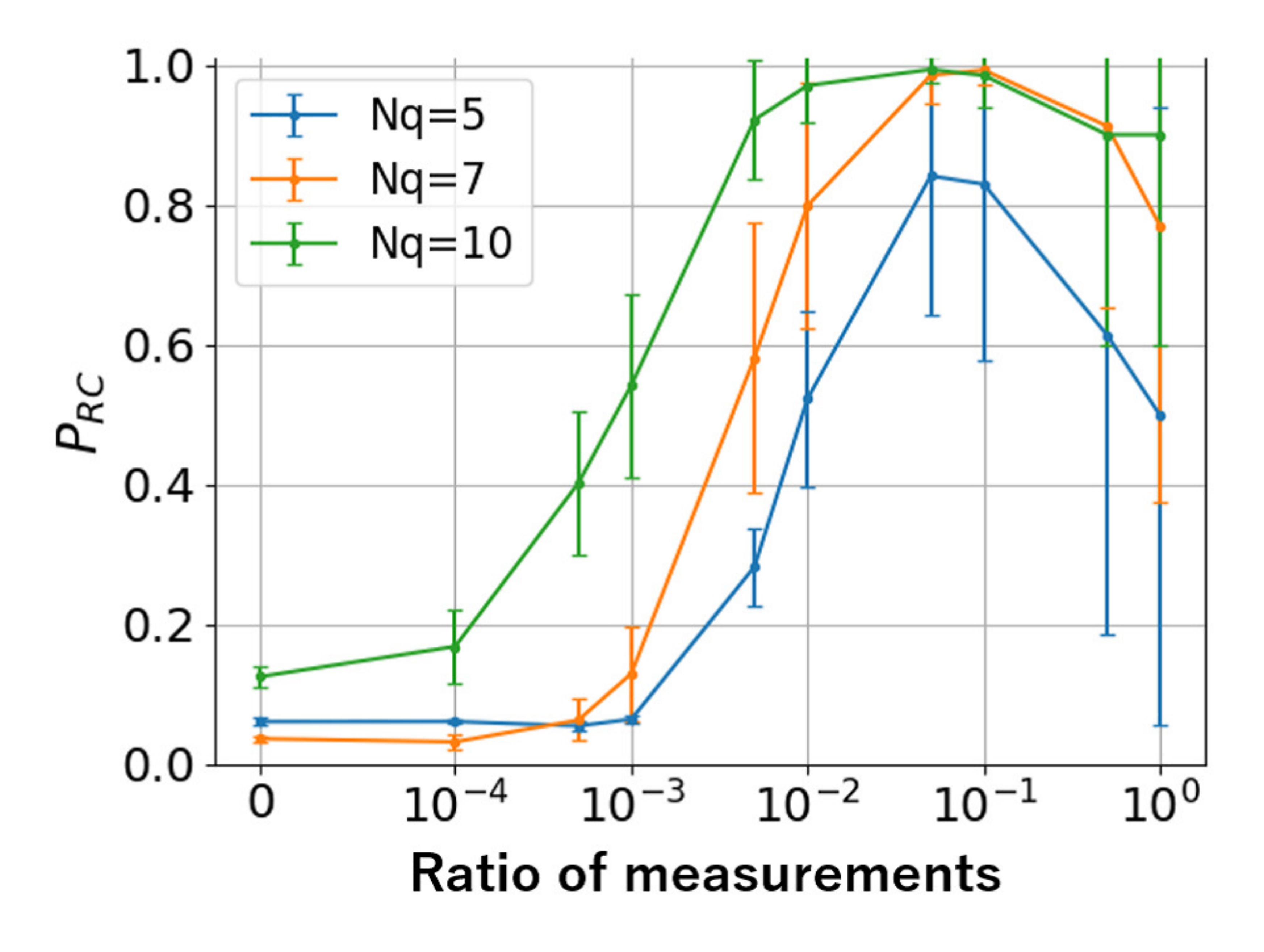}
  \end{minipage}
  \begin{minipage}[b]{0.4\textwidth}
    \centering
    \includegraphics[width=\textwidth]{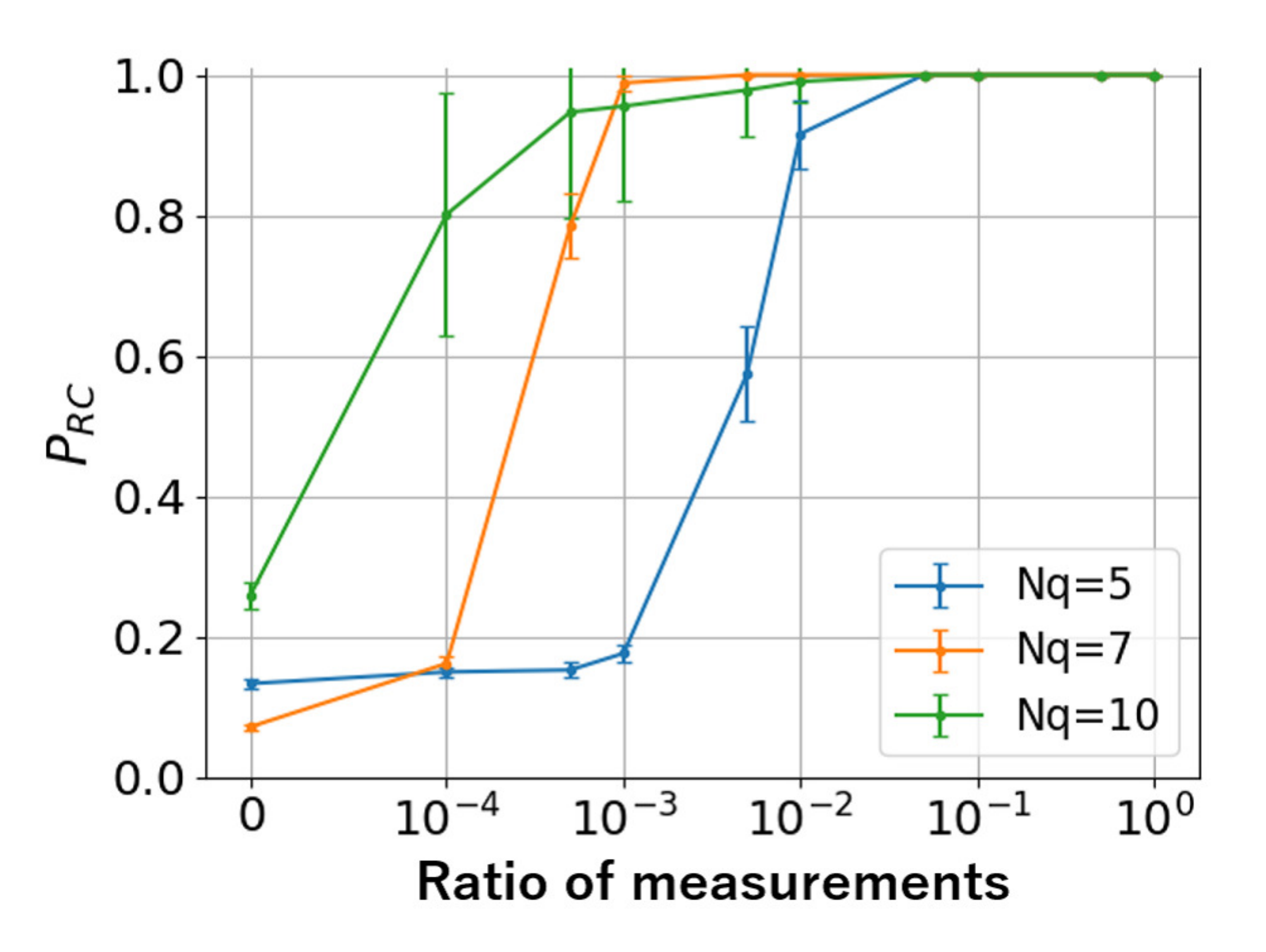}
  \end{minipage}
  \caption{Classification results of ${\sf noisy LRC}(D, p)$ for various system sizes $N_q$ and the measurement ratio $p$, where the depth $D$ is sufficiently large. The left- and right-hand sides correspond to the classification by the NN classifiers with $k'=4$ and $10$, respectively. The dot is the average and the error bar is the standard deviation over $10$ classifiers.}
  \label{pic_LRC_MIPT_check}
\end{figure*}

A toy model of noisy LRCs, ${\sf noisy LRC}(D, p)$, is depicted in FIG.~\ref{pic_circ_LRCwithNoise}, where $D$ is the depth and $p$ is the depolarizing parameter of the noise defined by Eq.~\eqref{def_DepolarizingChannel}. 
To see how ${\sf noisy LRC}(D, p)$ is classified, we apply the NN classifiers constructed for $k' = 4$ in the last section to the test data generated by a noisy ${\sf noisy LRC}(D, p)$ for various $D$ and $p$. The size of each test data set is chosen as $1000$. The results are provided in FIG.~\ref{pic_7q_application_s.o.r}.

As we have discussed in Subsec.~\ref{SS:LRCvarSum}, ${\sf noisy LRC}(D, p)$ with even tiny depolarizing parameter such as $p \leq 10^{-3}$ leads to a drastic change into the classification result, and a `dip' appears for non-zero $p$. This can be intuitively explained by observing how the data set changes with $p$ increases.
To clarify it, we visualize the datasets by the Principal Component Analysis (PCA), where we project the feature vectors of the data set of ${\sf noisy LRC}(D, p)$ into the $2$-dimensional subspace of the first and second principal components (PC1 and PC2) of the data set for ${\sf RC}$ and ${\sf Haar}$. 
The results of the PCA in the case of $k'=4$ is given in FIG.~\ref{pic_pca_noise_d} for $D \in \{10,14,18,24\}$.
Note that due to the z-score normalization, the distribution of the data set for ${\sf RC}$ and ${\sf Haar}$ centers around the origin.

A generic feature of the data set of ${\sf noisy LRC}(D, p)$ for $p\neq 0$ is as follows.
When $D$ is sufficiently small, such as $D=10$, it is far from those of ${\sf RC}$ and of ${\sf Haar}$ for any $p$ including $p=0$. As $D$ increases, it gradually approaches to that of ${\sf Haar}$. Then, for sufficiently large $D$, it moves away again.
The depth, at which the data set ${\sf noisy LRC}(D, p)$ gets the closest to that of ${\sf Haar}$, depends on the value of $p$: it is $\approx 18$ for $p = 5 \times 10^{-5}$ and $\approx 10$ for $p = 1 \times 10^{-3}$, which in fact coincides with the depth at which we observe a `dip' in FIG.~\ref{pic_7q_application_s.o.r}.

\subsection{Characterization of the monitored LRCs} \label{SS:moni}

We apply the NN classifier to the data set generated by the monitored LRCs with depth $D$ and measurement ratio $p$, ${\sf monitLRC}(D,p)$. See FIG.~\ref{pic_circ_LRCwithMeasurement} for the definition of the monitored LRCs.
The classification result of the data set generated by ${\sf monitLRC}(D,p)$ by the NN classifier is given in FIG.  \ref{pic_7q_application_s.o.r} for various $D$ and $p$.

As we have discussed in Subsec.~\ref{SS:LRCvarSum}, the classification result seems to fail to capture the MIPT, which is the prominent characteristic of ${\sf monitLRC}(D,p)$. 
To clarify this point, we plot the classification result $P_{RC}$ by NN classifiers for $k'=4$ and $10$ as a function of the measurement ratio $p$ of ${\sf monitLRC}(D,p)$ for various system sizes $N_q$, where $D$ is taken to be sufficiently large.
The results are given in FIG.~\ref{pic_LRC_MIPT_check}. For both $k' = 4$ and $10$, we observe that, as the system size $N_q$ increases, the classification result is more likely to be ${\sf RC}$ for any value of the measurement ratio $p$. This is clear especially for $k'=4$, where the results for $N_q=10$ indicate that the classification is ${\sf RC}$ even for tiny ratio of measurements. A similar behavior is expected for $k'=10$.
Thus, it is reasonable to expect that the classification of ${\sf monitLRC}(D,p)$ for $p\neq 0$ is always ${\sf RC}$ when the system size is sufficiently large, which indicates that the MIPT cannot observed by $P_{RC}$.

From this result, we may conjecture that MIPTs are not transitions of random dynamics in terms of $t$-designs. However, it is hard to conclude from our results since we have investigated only one quantity $P_{RC}$ related to $t$-designs. There may exist a quantity related to $t$-designs such that an MIPT can be observed in terms of the quantity. We hence leave this question, whether or not MIPTs are transitions related to $t$-designs, as a future problem.

Despite the fact that MIPTs are not observed from the classification result, the classification result of ${\sf monitLRC}(D,p)$ visibly differs from that of ${\sf noisyLRC}(D,p)$. Hence, the classifier provides a hint if experimentally-implemented monitored LRCs are noiseless or noisy and may help practical realizations of MIPTs.

\section{Conclusions} \label{sec of conclusion}

We have proposed a new supervised learning approach for characterizing random dynamics in terms of a $t$-design. Our method uses the data set experimentally easy-to-access, namely, measurement outcomes in the computational basis. By learning the moments of the multi-bit correlations of the data generated by ${\sf RC}$ and ${\sf Haar}$, which are estimated from a finite number of measurements, we have constructed classifiers that outputs either ${\sf RC}$ or ${\sf Haar}$. We have tries several learning algorithms, such as LR, LSVM, SVM, RF, and NN, and have shown that SVM and NN have good performance, which imply that it is possible to characterize ${\sf RC}$ and ${\sf Haar}$ only from the measurement outcomes.

We have then applied the NN classifier to the data set generated by LRCs, which exhibit growth of circuit complexity in the sense of a $t$-design as the depth grows. Despite the fact that the data set is completely different from those generated by ${\sf RC}$ and ${\sf Haar}$, it has turned out that the classifier successfully detects the transition of LRCs. Thus, the NN classifier is capable to capture the properties of random dynamics related to $t$-designs, not those specific to ${\sf RC}$ and ${\sf Haar}$. This is essential if one applies the classifier to experimentally implemented random dynamics since the details of the dynamics should be in general unknown.

We have finally applied the NN classifier to noisy and monitored LRCs to check its wide applicability. In the former case, we have found that the output of the NN classifier is sensitive to the presence of noise in LRCs, opening the possibility of using the classifier for verification of noisy quantum devices. In the latter case, the classifier has shown to fail detecting the MIPTs, a prominent feature of the monitored LRCs. This possibly indicates that MIPTs may not be transitions in terms of $t$-designs, though a detailed analysis is open.
We, however, expect that the NN classifier helps experimental realizations of monitored LRCs since MIPTs are unlikely to be robust against noise and the classifier is sensitive to the presence of noise.

This paper has established that supervised learning methods are useful for verifying unitary $t$-designs from an experimentally easy-to-access data, but there are a number of open questions. From a technical viewpoint, it would be the most important to decide the best moment $k'$ for distinguishing a $t$-design from higher-designs. As mentioned, despite the fact that it is theoretically natural to choose $k'=t$, our numerical results indicate that choosing larger $k'$ seem to result in better performance. It is open if this is always true and also if setting $k'$ as large as possible is a better strategy. Addressing this both numerically and theoretically is an important problem.

Another interesting direction is to explore different methods to characterize $t$-designs based on experimentally easy-to-access data. A possible way is to use the Porter-Thomas distribution, which is used for verification of noisy random circuits~\cite{B2018} and also for characterizing $t$-designs~\cite{Iaconis2021}. 
Despite the facts that the relation between the Porter-Thomas distribution and designs are not clarified yet and that how the Porter-Thomas distribution changes in the presence of statistical errors caused by a finite number of measurements, the analysis based on the distribution will lead to another necessary condition for the dynamics being a $t$-design. By combining the condition with the one obtained from our analysis, a better understanding of random dynamics in terms of designs will be obtained.

From the viewpoint of applications, $t$-designs have a huge number of applications from quantum information to theoretical physics, in both of which experimental realizations are the final goal. Hence, verification of experimental realizations should one of the most significant problems in both fields. It is expected that our approach, using supervised learning, is useful for this task in general. In this paper, we have verified quantum pseudorandomness, i.e., $t$-designs, but the method should work for direct verification of general tasks. Thus, applying the supervised learning method directly to verifying experiments related to quantum pseudorandomness is an interesting and important future direction.

\section*{Code availability}
The source code for generating datasets and classifiers used in this paper is available in \footnote{Source code is available here: \url{https://github.com/mf-22/t-design_MPbased_cpp}}.

\section*{Acknowledgements}
In this research work, we used the supercomputer of ACCMS, Kyoto University.
Y. Nakata is supported by JST, PRESTO Grant Number JPMJPR1865, Japan, and partially by MEXT-JSPS Grant-in-Aid for Transformative Research Areas (A) ”Extreme Universe”, Grant Numbers JP21H05182 and JP21H05183, and by JSPS KAKENHI Grant Number JP22K03464.
Y. Suzuki is supported by JST PRESTO Grant Number JPMJPR1916; MEXT Q-LEAP Grant Number JPMXS0120319794 and JPMXS0118068682; Moonshot R\&D, JST, Grant Number JPMJMS2061.
M.Owari is supported by JSPS KAKENHI Grant Numbers JP21K03388, JP20K03779.

\bibliographystyle{unsrt}
\bibliography{myref}% Produces the bibliography via BibTeX.

\providecommand{\noopsort}[1]{}\providecommand{\singleletter}[1]{#1}%
\begin{thebibliography}{10}

\bibitem{M2014RM}
M.~L. Mehta.
\newblock {\em Random Matrices (Pure and Applied Mathematics)}.
\newblock Academic Press, 2014.

\bibitem{AE2007}
A.~Ambainis and J.~Emerson.
\newblock Quantum $t$-designs: $t$-wise independence in the quantum world.
\newblock In {\em Proc. IEEE Conf. Comput. 2007}, page 129–140, 2007.
\newblock arXiv:quant-ph/0701126v2.

\bibitem{L2010}
R.~A. Low.
\newblock {\em Pseudo-randomness and learning in quantum computation}.
\newblock PhD thesis, University of Bristol, 2010.
\newblock arXiv:1006.5227.

\bibitem{D2005}
I.~Devetak.
\newblock The private classical capacity and quantum capacity of a quantum
  channel.
\newblock {\em IEEE Trans. Inf. Theory}, 51:44--55, 2005.

\bibitem{DW2004}
I.~Devetak and A.~Winter.
\newblock Relating {Quantum} {Privacy} and {Quantum} {Coherence}: {An}
  {Operational} {Approach}.
\newblock {\em Phys. Rev. Lett.}, 93:080501, 2004.

\bibitem{GPW2005}
B.~Groisman, S.~Popescu, and A.~Winter.
\newblock Quantum, classical, and total amount of correlations in a quantum
  state.
\newblock {\em Phys. Rev. A}, 72:032317, 2005.

\bibitem{ADHW2009}
A.~Abeyesinghe, I.~Devetak, P.~Hayden, and A.~Winter.
\newblock The mother of all protocols : Restructuring quantum information's
  family tree.
\newblock {\em Proc. R. Soc. A}, 465:2537--2563, 2009.

\bibitem{DBWR2010}
F.~Dupuis, M.~Berta, J.~Wullschleger, and R.~Renner.
\newblock One-shot decoupling.
\newblock {\em Commun. Math. Phys.}, 328:251--284, 2014.

\bibitem{SDTR2013}
O.~Szehr, F.~Dupuis, M.~Tomamichel, and R.~Renner.
\newblock Decoupling with unitary approximate two-designs.
\newblock {\em New J. Phys.}, 15:053022, 2013.

\bibitem{HOW2005}
M.~Horodecki, J.~Oppenheim, and A.~Winter.
\newblock Partial quantum information.
\newblock {\em Nature}, 436:673--676, 2005.

\bibitem{HOW2007}
M.~Horodecki, J.~Oppenheim, and A.~Winter.
\newblock Quantum state merging and negative information.
\newblock {\em Comms. Math. Phys.}, 269:107--136, 2007.

\bibitem{NWY2021}
Y.~Nakata, E.~Wakakuwa, and H.~Yamasaki.
\newblock One-shot quantum error correction of classical and quantum
  information.
\newblock {\em Phys. Rev. A}, 104:012408, 2021.

\bibitem{AS2004}
A.~Ambainis and A.~Smith.
\newblock Small pseudo-random families of matrices: Derandomizing approximate
  quantum encryption.
\newblock In {\em Proc. RANDOM'04}, page 249, 2004.

\bibitem{HLSW2004}
P.~Hayden, D.~Leung, P.~W. Shor, and A.~Winter.
\newblock Randomizing quantum states: Constructions and applications.
\newblock {\em Commun. Math. Phys.}, 250:371--391, 2004.

\bibitem{S2005}
P.~{Sen}.
\newblock Random measurement bases, quantum state distinction and applications
  to the hidden subgroup problem.
\newblock In {\em 21st Annual IEEE Conference on Computational Complexity
  (CCC'06)}, pages 14--287, 2006.

\bibitem{BH2013}
F.~G. S.~L. Brand{\~a}o and M.~Horodecki.
\newblock Exponential {Quantum} {Speed}-ups are {Generic}.
\newblock {\em Q. Inf. Comp.}, 13:0901, 2013.

\bibitem{KRT2014}
R.~Kueng, H.~Rauhut, and U.~Terstiege.
\newblock Low rank matrix recovery from rank one measurements.
\newblock {\em Appl. Comput. Harmon. Anal.}, 42:88--116, 2017.

\bibitem{KL15}
S.~{Kimmel} and Y.~{Liu}.
\newblock Phase retrieval using unitary 2-designs.
\newblock In {\em 2017 International Conference on Sampling Theory and
  Applications (SampTA)}, pages 345--349, 2017.

\bibitem{KZD2016}
R.~Kueng, H.~Zhu, and D.~Gross.
\newblock Distinguishing quantum states using {Clifford} orbits, 2016.
\newblock arXiv:1609.08595.

\bibitem{OAGKAL2016}
M.~Oszmaniec, R.~Augusiak, C.~Gogolin, J.~Ko\l{}ody\'{n}ski, A.~Ac\'{i}n, and
  M.~Lewenstein.
\newblock Random {Bosonic} {States} for {Robust} {Quantum} {Metrology}.
\newblock {\em Phys. Rev. X}, 6:041044, 2016.

\bibitem{EAZ2005}
J.~Emerson, R.~Alicki, and K.~{\.Z}yczkowski.
\newblock Scalable noise estimation with random unitary operators.
\newblock {\em J. Opt. B: Quantum semiclass. opt.}, 7:S347--S352, 2005.

\bibitem{KLRetc2008}
E.~Knill et~al.
\newblock Randomized benchmarking of quantum gates.
\newblock {\em Phys. Rev. A}, 77:012307, 2008.

\bibitem{MGE2011}
E.~Magesan, J.~M. Gambetta, and J.~Emerson.
\newblock Scalable and {Robust} {Randomized} {Benchmarking} of {Quantum}
  {Processes}.
\newblock {\em Phys. Rev. Lett.}, 106:180504, 2011.

\bibitem{MGE2012}
E.~Magesan, J.~M. Gambetta, and J.~Emerson.
\newblock Characterizing quantum gates via randomized benchmarking.
\newblock {\em Phys. Rev. A}, 85:042311, 2012.

\bibitem{B2018}
S.~Boixo et~al.
\newblock Characterizing quantum supremacy in near-term devices.
\newblock {\em Nat. Phys.}, 14:595--600, 2018.

\bibitem{G2019}
F~Arute et~al.
\newblock Quantum supremacy using a programmable superconducting processor.
\newblock {\em Nature}, 574:505--510, 2019.

\bibitem{BFNV2019}
A.~Bouland, B.~Fefferman, C.~Nirkhe, and U.~Vazirani.
\newblock On the complexity and verification of quantum random circuit
  sampling.
\newblock {\em Nat. Phys.}, 15:159--163, 2019.

\bibitem{PSW2006}
S.~Popescu, A.~J. Short, and A.~Winter.
\newblock Entanglement and the foundations of statistical mechanics.
\newblock {\em Nat. Phys.}, 2:754--758, 2006.

\bibitem{dRARDV2011}
L.~del Rio, J.~Aberg, R.~Renner, O.~Dahlsten, and V.~Vedral.
\newblock The thermodynamic meaning of negative entropy.
\newblock {\em Nature}, 474:61--63, 2011.

\bibitem{dRHRW2016}
L.~del Rio, A.~Hutter, R.~Renner, and S.~Wehner.
\newblock Relative thermalization.
\newblock {\em Phys. Rev. E}, 94:022104, 2016.

\bibitem{HP2007}
P.~Hayden and J.~Preskill.
\newblock Black holes as mirrors: quantum information in random subsystems.
\newblock {\em J. High Energy Phys.}, 2007:120, 2007.

\bibitem{SS2008}
Y.~Sekino and L.~Susskind.
\newblock Fast scramblers.
\newblock {\em J. High Energy Phys.}, 2008:065, 2008.

\bibitem{S2011}
L.~Susskind.
\newblock Addendum to {Fast} {Scramblers}.
\newblock {\em arXiv:1101.6048}, 2011.

\bibitem{LSHOH2013}
N.~Lashkari, D.~Stanford, M.~Hastings, T.~Osborne, and P.~Hayden.
\newblock Towards the fast scrambling conjecture.
\newblock {\em J. High Energy Phys.}, 2013:22, 2013.

\bibitem{HQRY2016}
P.~Hosur, X.-L. Qi, D.~A. Roberts, and B.~Yoshida.
\newblock Chaos in quantum channels.
\newblock {\em J. High Energy Phys.}, 2016:4, 2016.

\bibitem{RY2017}
D.~A. Roberts and B.~Yoshida.
\newblock Chaos and complexity by design.
\newblock {\em J. High Energ. Phys.}, 2017:121, 2017.

\bibitem{NWK2020}
Y.~Nakata, E.~Wakakuwa, and M.~Koashi.
\newblock Information leakage from quantum black holes with symmetry.
\newblock {\em arXiv:2007.00895}, 2020.

\bibitem{LFSLYYM2019}
K.~A. Landsman et~al.
\newblock Verified quantum information scrambling.
\newblock {\em Nature}, 567:61--65, 2019.

\bibitem{LCF2018}
Y.~Li, X.~Chen, and M.~P.~A. Fisher.
\newblock Quantum zeno effect and the many-body entanglement transition.
\newblock {\em Phys. Rev. B}, 98:205136, 2018.

\bibitem{JYVL2020}
C.~Jian, Y.~You, R.~Vasseur, and A.~W.~W. Ludwig.
\newblock Measurement-induced criticality in random quantum circuits.
\newblock {\em Phys. Rev. B}, 101:104302, 2020.

\bibitem{BCA2020}
Y.~Bao, S.~Choi, and E.~Altman.
\newblock Theory of the phase transition in random unitary circuits with
  measurements.
\newblock {\em Phys. Rev. B}, 101:104301, 2020.

\bibitem{HL2009TPE}
A.~W. Harrow and R.~A. Low.
\newblock Efficient {Quantum} {Tensor} {Product} {Expanders} and k-{Designs}.
\newblock In {\em Proc. RANDOM'09}, Lecture {Notes} in {Computer} {Science},
  pages 548--561. Springer Berlin Heidelberg, 2009.

\bibitem{BHH2016}
F.~G. S.~L. Brand{\~a}o, A.~W. Harrow, and M.~Horodecki.
\newblock Local {Random} {Quantum} {Circuits} are {Approximate}
  {Polynomial}-{Designs}.
\newblock {\em Commun. Math. Phys.}, 346:397--434, 2016.

\bibitem{HM2018}
A.~Harrow and S.~Mehraban.
\newblock Approximate unitary $t$-designs by short random quantum circuits
  using nearest-neighbor and long-range gates.
\newblock {\em arXiv:1809.06957}, 2018.

\bibitem{HMHEGR2020}
J.~Haferkamp et~al.
\newblock Quantum homeopathy works: {Efficient} unitary designs with a
  system-size independent number of non-{Clifford} gates.
\newblock {\em arXiv:2002.09524}, 2020.

\bibitem{NZOBSTHYZTTN2021}
Y.~Nakata et~al.
\newblock Quantum circuits for exact unitary $t$-designs and applications to
  higher-order randomized benchmarking.
\newblock {\em PRX Quantum}, 2:030339, 2021.

\bibitem{NHKW2017}
Y.~Nakata, C.~Hirche, M.~Koashi, and A.~Winter.
\newblock Efficient {Quantum} {Pseudorandomness} with {Nearly}
  {Time}-{Independent} {Hamiltonian} {Dynamics}.
\newblock {\em Phys. Rev. X}, 7:021006, 2017.

\bibitem{Lietal2019}
J.~Li et~al.
\newblock Experimental implementation of efficient quantum pseudorandomness on
  a 12-spin system.
\newblock {\em Phys. Rev. Lett.}, 123:030502, 2019.

\bibitem{Scott_2008}
A~J Scott.
\newblock Optimizing quantum process tomography with unitary 2-designs.
\newblock {\em Journal of Physics A: Mathematical and Theoretical},
  41(5):055308, jan 2008.

\bibitem{LO1969}
A.~Larkin and Y.~N. Ovchinnikov.
\newblock Quasiclassical method in the theory of superconductivity.
\newblock {\em Sov. Phys. JETP}, 28:1200, 1969.

\bibitem{Almheirietal2013}
A.~Almheiri et~al.
\newblock An apologia for firewalls.
\newblock {\em J. High Energ. Phys.}, page~18, 2013.

\bibitem{Shenker2014}
Stephen~H. Shenker and Douglas Stanford.
\newblock Black holes and the butterfly effect.
\newblock {\em Journal of High Energy Physics}, 2014(3):67, Mar 2014.

\bibitem{Shaffer_2014}
Daniel Shaffer, Claudio Chamon, Alioscia Hamma, and Eduardo~R Mucciolo.
\newblock Irreversibility and entanglement spectrum statistics in quantum
  circuits.
\newblock {\em Journal of Statistical Mechanics: Theory and Experiment},
  2014(12):P12007, dec 2014.

\bibitem{CHM2014}
Claudio Chamon, Alioscia Hamma, and Eduardo~R. Mucciolo.
\newblock Emergent irreversibility and entanglement spectrum statistics.
\newblock {\em Phys. Rev. Lett.}, 112:240501, Jun 2014.

\bibitem{YHGMC2017}
Zhi-Cheng Yang, Alioscia Hamma, Salvatore~M. Giampaolo, Eduardo~R. Mucciolo,
  and Claudio Chamon.
\newblock Entanglement complexity in quantum many-body dynamics,
  thermalization, and localization.
\newblock {\em Phys. Rev. B}, 96:020408, Jul 2017.

\bibitem{Liu2018}
Zi-Wen Liu, Seth Lloyd, Elton Zhu, and Huangjun Zhu.
\newblock Entanglement, quantum randomness, and complexity beyond scrambling.
\newblock {\em Journal of High Energy Physics}, 2018(7):41, Jul 2018.

\bibitem{BCHKP21}
Fernando~G.S.L. Brand\~ao, Wissam Chemissany, Nicholas Hunter-Jones, Richard
  Kueng, and John Preskill.
\newblock Models of quantum complexity growth.
\newblock {\em PRX Quantum}, 2:030316, Jul 2021.

\bibitem{Iaconis2021}
Jason Iaconis.
\newblock Quantum state complexity in computationally tractable quantum
  circuits.
\newblock {\em PRX Quantum}, 2:010329, Feb 2021.

\bibitem{PhysRevB.94.165134}
G.~Torlai and R.~G. Melko.
\newblock Learning thermodynamics with boltzmann machines.
\newblock {\em Phys. Rev. B}, 94:165134, Oct 2016.

\bibitem{doi:10.1126/science.aag2302}
G.~Carleo and M.~Troyer.
\newblock Solving the quantum many-body problem with artificial neural
  networks.
\newblock {\em Science}, 355(6325):602--606, 2017.

\bibitem{torlai_neural-network_2018}
G.~Torlai, G.~Mazzola, J.~Carrasquilla, M.~Troyer, R.~Melko, and G.~Carleo.
\newblock Neural-network quantum state tomography.
\newblock {\em Nat. Phys.}, 14:447--450, 2018.

\bibitem{PhysRevX.8.021050}
M.~H. Amin, E.~Andriyash, J.~Rolfe, B.~Kulchytskyy, and R.~Melko.
\newblock Quantum boltzmann machine.
\newblock {\em Phys. Rev. X}, 8:021050, May 2018.

\bibitem{carrasquilla_reconstructing_2019}
J.~Carrasquilla, G.~Torlai, R.~G. Melko, and L.~Aolita.
\newblock Reconstructing quantum states with generative models.
\newblock {\em Nat. Mach. Intell}, 1(3):155--161, 2019.

\bibitem{doi:10.1146/annurev-conmatphys-031119-050651}
G.~Torlai and R.~G. Melko.
\newblock Machine-learning quantum states in the nisq era.
\newblock {\em Annu. Rev. Condens. Matter Phys.}, 11(1):325--344, 2020.

\bibitem{alves2020machine}
D.~W.~F. Alves and M.O.Flynn.
\newblock Machine learning, quantum chaos, and pseudorandom evolution.
\newblock {\em Phys. Rev. A}, 101:052338, 2020.

\bibitem{Zhu2017}
H.~Zhu.
\newblock Multiqubit clifford groups are unitary 3-designs.
\newblock {\em Phys. Rev. A}, 96:062336, 2017.

\bibitem{W2016}
Z.~Webb.
\newblock The {Clifford} group forms a unitary 3-design.
\newblock {\em Quant. Info. {\&} Comp.}, 16:1379--1400, 2016.

\bibitem{Haferkamp2022}
Jonas Haferkamp, Philippe Faist, Naga B.~T. Kothakonda, Jens Eisert, and Nicole
  Yunger~Halpern.
\newblock Linear growth of quantum circuit complexity.
\newblock {\em Nature Physics}, 18(5):528--532, May 2022.

\bibitem{emerson2003pseudo}
J.~Emerson ant others.
\newblock Pseudo-random unitary operators for quantum information processing.
\newblock {\em science}, 302(5653):2098--2100, 2003.

\bibitem{PQSV11}
David Poulin, Angie Qarry, Rolando Somma, and Frank Verstraete.
\newblock Quantum simulation of time-dependent hamiltonians and the convenient
  illusion of hilbert space.
\newblock {\em Phys. Rev. Lett.}, 106:170501, Apr 2011.

\bibitem{KSV2002}
A.~Kitaev, A.~Shen, and M.~Vyalyi.
\newblock {\em {\it Classical and Quantum Computation}}.
\newblock American Mathematical Society Boston, MA, USA, 2002.

\bibitem{BNOZ2020}
E.~Bannai, Y.~Nakata, T.~Okuda, and D.~Zhao.
\newblock Explicit construction of exact unitary designs.
\newblock {\em accepted in Adv. Math. (see also arXiv:2009.11170)}, 2020.

\bibitem{GSW2021}
M.~A. Graydon, J.~Skanes-Norman, and J.~J. Wallman.
\newblock Clifford groups are not always 2-designs.
\newblock {\em arXiv:2108.04200}, 2021.

\bibitem{HJ2019}
N.~Hunter-Jones.
\newblock Unitary designs from statistical mechanics in random quantum
  circuits.
\newblock {\em arXiv:1905.12053}, 2019.

\bibitem{Li2017measuring}
J.Li, R.Fan, H.Wang, B.Ye amd B.Zeng, H.Zhai, X.Peng, and J.Du.
\newblock Measuring out-of-time-order correlators on a nuclear magnetic
  resonance quantum simulator.
\newblock {\em Phys. Rev. X}, 7:031011, Jul 2017.

\bibitem{mi2021information}
et~al. X.Mi.
\newblock Information scrambling in quantum circuits.
\newblock {\em Science}, 374(6574):1479--1483, 2021.

\bibitem{PhysRevB.101.104301}
Yimu Bao, Soonwon Choi, and Ehud Altman.
\newblock Theory of the phase transition in random unitary circuits with
  measurements.
\newblock {\em Phys. Rev. B}, 101:104301, Mar 2020.

\bibitem{PhysRevX.10.041020}
Michael~J. Gullans and David~A. Huse.
\newblock Dynamical purification phase transition induced by quantum
  measurements.
\newblock {\em Phys. Rev. X}, 10:041020, Oct 2020.

\bibitem{PhysRevB.101.104302}
Chao-Ming Jian, Yi-Zhuang You, Romain Vasseur, and Andreas W.~W. Ludwig.
\newblock Measurement-induced criticality in random quantum circuits.
\newblock {\em Phys. Rev. B}, 101:104302, Mar 2020.

\bibitem{suzuki2020qulacs}
Y.~Suzuki et~al.
\newblock Qulacs: a fast and versatile quantum circuit simulator for research
  purpose.
\newblock {\em Quantum}, 5:559, 2021.

\bibitem{lmcs:1570}
P.~Selinger.
\newblock {Generators and relations for n-qubit Clifford operators}.
\newblock {\em {Logical Methods in Computer Science}}, {Volume 11, Issue 2},
  2015.

\bibitem{chollet2015keras}
Fran\c{c}ois Chollet et~al.
\newblock Keras.
\newblock \url{https://keras.io}, 2015.

\bibitem{tensorflow2015-whitepaper}
Mart\'{\i}n Abadi et~al.
\newblock {TensorFlow}: Large-scale machine learning on heterogeneous systems,
  2015.
\newblock Software available from tensorflow.org.

\bibitem{scikit-learn}
F.~Pedregosa et~al.
\newblock Scikit-learn: Machine learning in {P}ython.
\newblock {\em Journal of Machine Learning Research}, 12:2825--2830, 2011.

\bibitem{Note1}
Source code is available here: \protect \url
  {https://github.com/mf-22/t-design_MPbased_cpp}.

\bibitem{L2001}
M.~Ledoux.
\newblock {\em The Concentration of Measure Phenomenon}.
\newblock American Mathematical Society Providence, RI, USA, 2001.

\end{thebibliography}

\appendix

\section{OTOC-based characterization} \label{sec of add. exp. of alves's method}

In~\cite{alves2020machine}, a classifier was constructed based on the data set of Out-Of-Time-Ordered Correlators (OTOCs), which are defined for an ensemble ${\sf W}$ of unitaries by 
\begin{equation}\label{OTOC random 2 i j}
O_{i,j}^{(2)}=   \mathbb{E}_{U \sim {\sf W}} \left [ {\rm Tr} \left( A_i U^\dagger B_j U A_i U^\dagger B_j U \right)\right],
\end{equation}
where $A_i$ and $B_j$ are the Pauli operators on the $i$th and $j$th qubits.
It was shown that, by learning such data sets, it is possible to construct a classifier that can characterize ${\sf Haar}$, the Pauli group ${\sf P}$, and a $2$-design constructed by RDCs.
Here, we reexamine their method, and a slightly more generalized one, and investigate if the classifier can also characterize the growth of circuit complexity in LRCs.

\subsection{Generating a data set}

Since the original OTOCs defined by Eq.~\eqref{OTOC random 2 i j} is difficult to measure in practice, it is necessary to consider variants of the original one. Here, we consider the original OTOCs as well as two variants:
\begin{align}
&O_{alv,i,j}:= \frac{1}{m} \sum_{n=1}^{m}\Bra{\sigma_n}A_iU^{\dag}_nB_jU_nA_iU^{\dag}_nB_jU_n\Ket{\sigma_n}, \label{otoc}\\
&O_{cb,i,j}:= \frac{1}{m} \sum_{n=1}^{m}\langle \vec{i}_n | A_iU^{\dag}_nB_jU_nA_iU^{\dag}_nB_jU_n |\vec{i}_n \rangle,\\
&O_{tr, i, j} := \frac{1}{m}\sum_{n=1}^{m} \operatorname{Tr}\left( A{ }_{i}^{\dagger} U_{n}^{\dagger} B_{j} U_{n} A_{i} U{ }_{n}^{\dagger} B_{j} U_{n}\right ), \label{tr} 
\end{align}
where $U_1, \dots, U_m$ are $m$ unitaries independently sampled from the ensemble ${\sf W}$ of unitaries, each $\left |\sigma_n \right \rangle$ is a random state with respect to ${\sf Haar}$, and $|\vec{i}_n\rangle$ is a computational-basis state ($\vec{i}_n\in \{0,1\}^{N_q}$).
We also fix the Pauli operators $A_i$ and $B_j$, such as $A_i=X_i$ and $B_j=Y_j$.

The first one is the quantity used in~\cite{alves2020machine}, which is mainly used for reproducing the result in the paper.
The second one is the quantity by replacing the random states in $O_{alv,i,j}$ with a random computational-basis state. This variant is important from the viewpoint of implementations since random states with respect to ${\sf Haar}$ are extremely hard to generate. The last one is a natural candidate as an estimator of OTOCs from a finite number of samples of unitaries. 

All of the above quantities are considered to form a $N_q \times N_q$ complex matrix with the indices $i$ and $j$, which we denote by $O_{alv}, O_{cb}$, and $O_{tr}$. These are the input of machine learning.
Thus, each feature vector consists of $N_q^2$ features.

\subsection{Machine learning of OTOCs, and classification of LRCs}

\begin{figure*}[tbh!]
\centering
   \begin{minipage}[tb]{0.32\textwidth}
  \includegraphics[width=\textwidth]{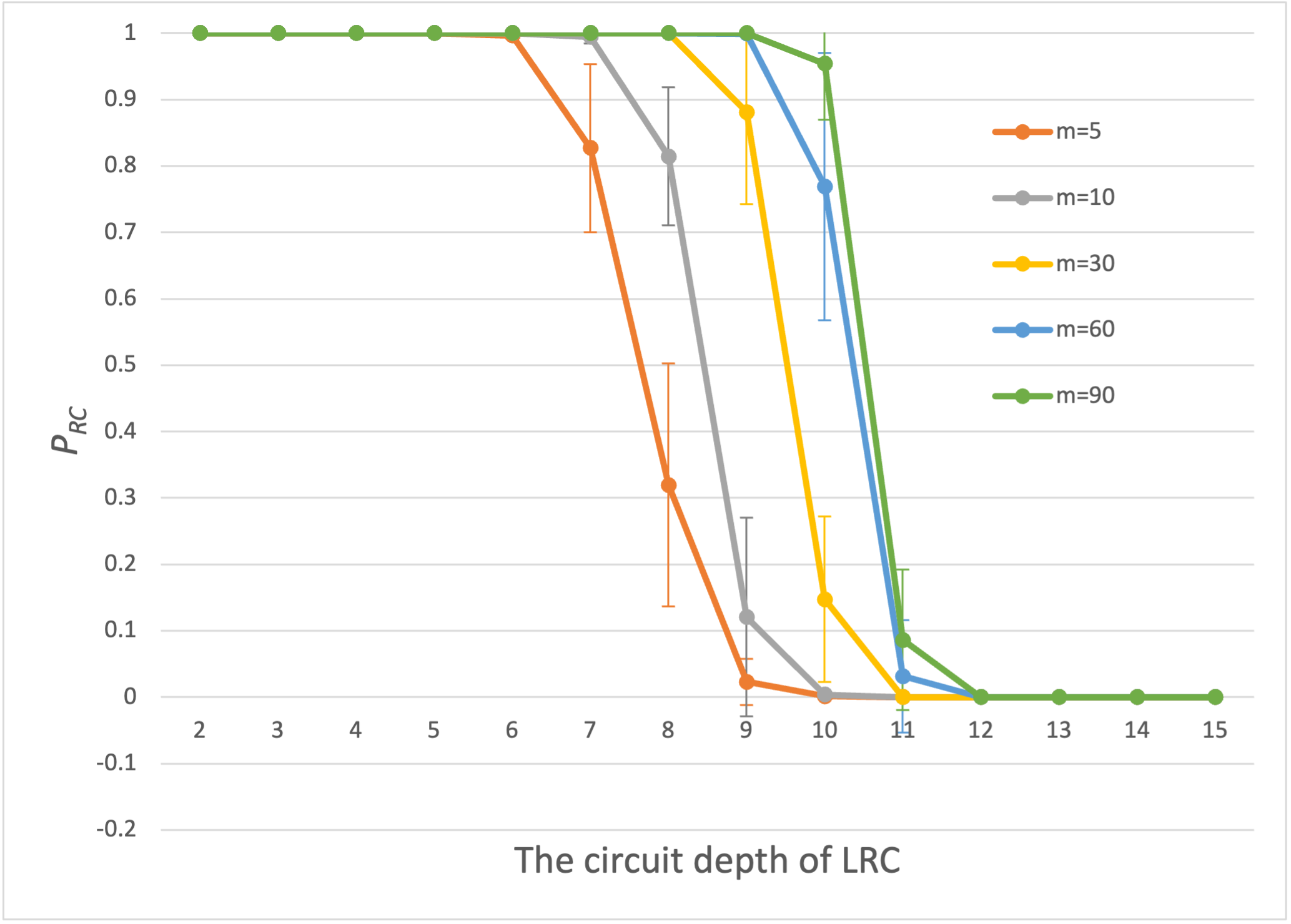}
  \end{minipage}
  \begin{minipage}[tb]{0.32\textwidth}
  \includegraphics[width=\textwidth]{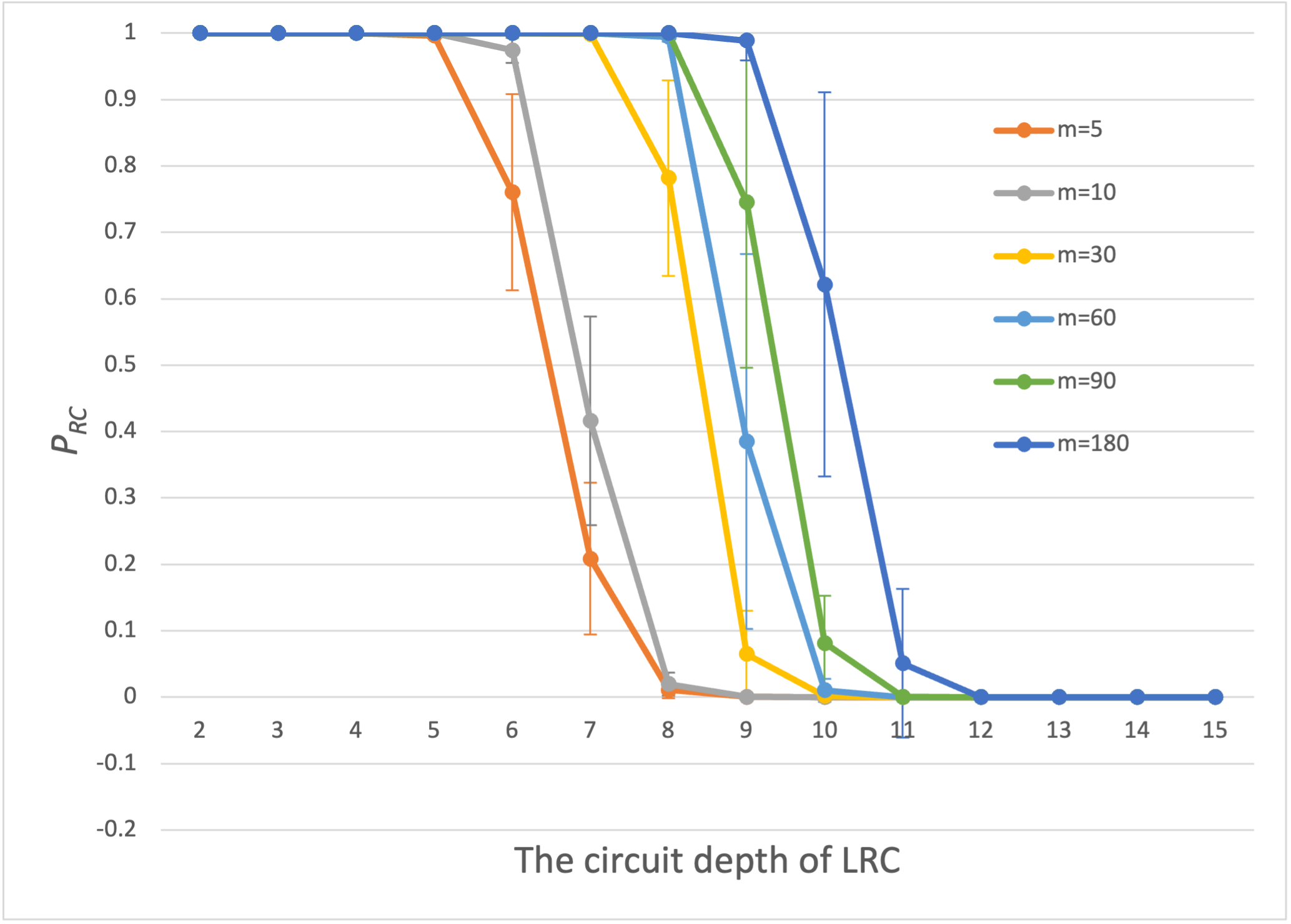}
  \end{minipage}
  \begin{minipage}[tb]{0.32\textwidth}
    \includegraphics[width=\textwidth]{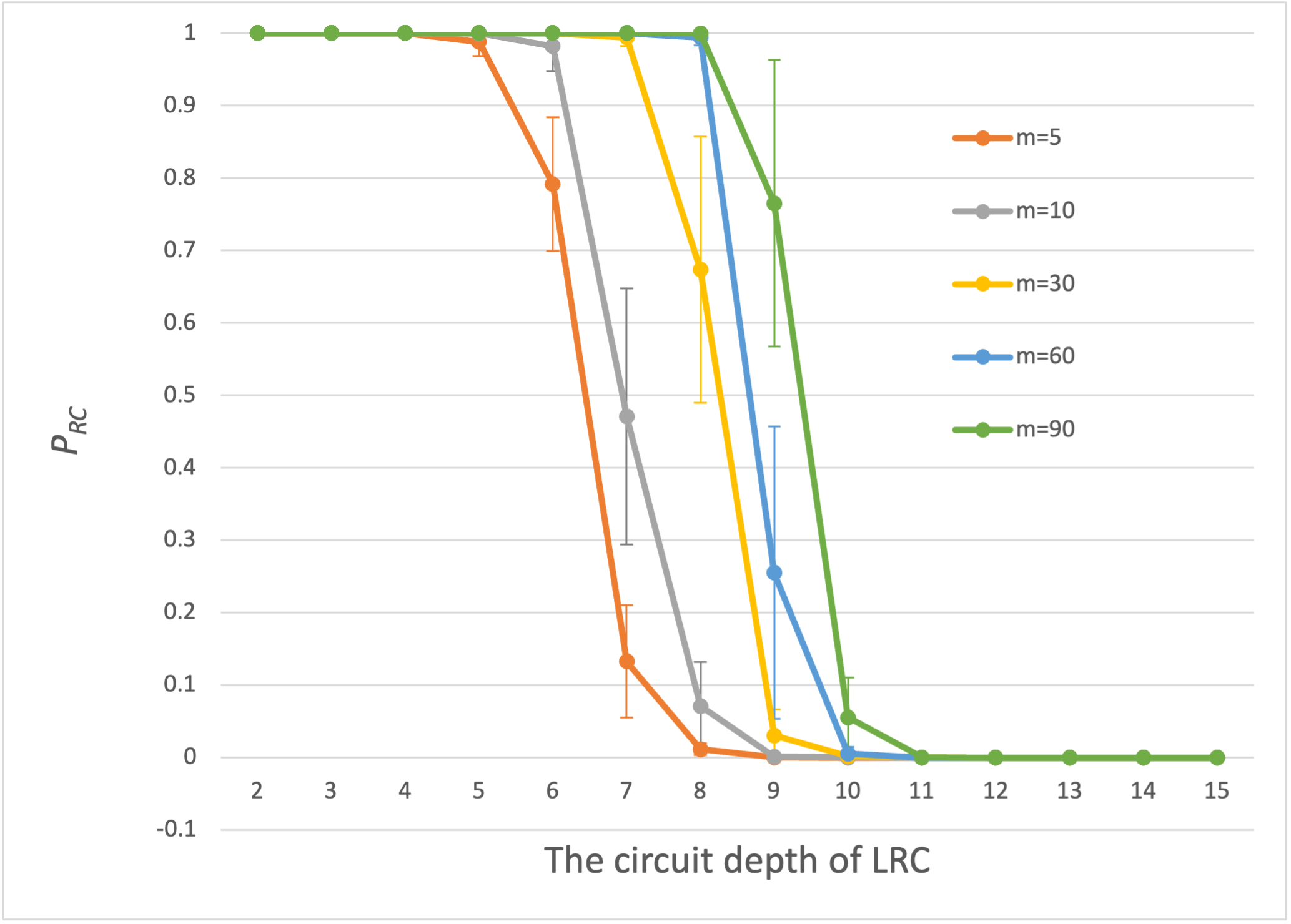}
  \end{minipage}
    \caption{Classification results of ${\sf LRC}(D)$ by the classifiers based on OTOCs for $5$ qubits. The classifiers are trained by NN methods from the data sets of $O_{tr}$, $O_{alve}$, and $O_{cb}$ from left to right, for various $m$. The vertical axis represents the probability $P_{RC}$ with which the classifier outputs ${\sf RC}$. The dots and the error bars are the averages and the standard deviations over $10$ classifiers constructed from the same but shuffled data set.}
  \label{lrc_alves_div}
\end{figure*}

We generate the data set, $O_{alv}, O_{cb}$, and $O_{tr}$, for ${\sf RC}$ and for ${\sf Haar}$, and construct classifiers.
Before we explain the learning methods, it is important to note that $m$ plays an important role. By taking the limit of $m \rightarrow \infty$, each value converges to the average over a given ensemble. However, since OTOCs are polynomials of unitaries of degree $2$, the average over ${\sf RC}$ is exactly equal to that over ${\sf Haar}$:
\begin{equation}
\mathbb{E}_{U \sim {\sf RC}}\bigl[ O_{x, i, j} \bigr] = \mathbb{E}_{U \sim {\sf Haar}}\bigl[ O_{x, i, j} \bigr],
\end{equation}
for all $x = alv, cb, tr$.
It is also straightforward to show, by using the unitary invariance of the Haar measure, that this quantity does not depend on $i$ and $j$.
Thus, what is learnt from the date sets generated by ${\sf RC}$ and ${\sf Haar}$ is higher-order statistical moments of the quantities, such as variance, which is only available when $m$ finite.

Following the method in~\cite{alves2020machine}, we use the data set $O_{alv}$ and train the classifier by using convolutional neural networks (CNNs). There, Domain Coloring is used to visualize the complex function by assigning colors on the complex plane. Then, a $N_q \times N_q$ complex matrix is converted to an $N_q \times N_q$-pixels image, which is learned by CNNs.
We also try another learning methods based on feedforward neural networks (NNs), where a matrix is converted into a $1$-dimensional real vector. In this case, we construct three classifiers, each from $O_{alv}, O_{cb}$, and $O_{tr}$.

For machine learning, we have used the Python library Keras~\cite{chollet2015keras} with Tensorflow~\cite{tensorflow2015-whitepaper} as the backend. 
In CNNs, we use the following hyperparameters: in the first convolutional layer (input layer), the number of filters is chosen to be 32, and the kernel size to be $2\times 2$. In the second convolutional layer, the number of filters is chosen to be 64, and the kernel size to be 3$\times$3. These two layers are followed by  a join layer having $512$ nodes and an output layer having a single node. 
We use a 3-layer NN with the following hyper parameters: The input and intermediate layers have $98$ nodes, and the output layer has a single node. 
For both CNN and NN, the activation function is set to a sigmoid function for the output layer and a relu function for the other layers. The error function is set to the binary cross entropy, and the learning rate is set to 0.001. The number of epochs and the batch size is set to 200 and 100, respectively.

In the case of CNNs, we find that the learning ${\sf RC}$ and ${\sf Haar}$ is successful for $5$ and $7$ qubits for all $m \in \{5,10,30,60,90 \}$. The accuracy of the validation and test data sets is both more than $99\%$ for all cases, which is consistent with~\cite{alves2020machine}. However, it fails to classify ${\sf LRC}(D)$: for any depth $D$, ${\sf LRC}(D)$ is classified to ${\sf Haar}$. This classification result does not capture the transition of ${\sf LRC}(D)$ and implies that the classifier has learnt properties that are specific to ${\sf RC}$ or to ${\sf Haar}$, rather than those directly related to $t$-designs.
Hence, we conclude that the classifiers constructed in this way fail to satisfy the second criterion presented in Subsec.~\ref{SS:TCC}.

In the case of learning with NNs, the accuracy of the validation and test data sets for all $m$ is also more than $99\%$. By applying the classifier to the data set generated by ${\sf LRC}(D)$ for $5$ qubits with various $m$, we obtain FIG.~\ref{lrc_alves_div}, where $P_{RC}$ is the probability that the classifier outputs ${\sf RC}$. Note that, since ${\sf LRC}(D)$ is not Clifford, the output of ${\sf RC}$ implies merely that the corresponding circuit has properties more similar to ${\sf RC}$ rather than those to ${\sf Haar}$. 

Although the classification results show transitions as the depth $D$ increases, the transition point highly depends on the value $m$, which makes it difficult to use the classifier for characterizing the complexity of ${\sf LRC}(D)$ in the sense of $t$-designs. For instance, it is not clear what $m$ we should use when we are interested in the transition from a $3$-design to a $4$-design.

We conclude the analysis of classifiers with OTOCs with a remark on the shifting of the transition points in FIG.~\ref{lrc_alves_div} as $m$ increases. In all figures of $O_{alv}, O_{cb}$, and $O_{tr}$, the transition points monotonically moves to right as $m$ increases. We may understand this shifting from the fact that the classifier is likely to characterize the data sets by the statistical moments higher than their averages, e.g, variances because the average for ${\sf RC}$ and that for ${\sf Haar}$ exactly coincide in the limit of $m \rightarrow \infty$ for all $i,j$, as mentioned in the beginning of this section. 
It is also known that the variance of ${\sf Haar}$ is extremely small, which is due to the so-called \emph{concentration of measure}~\cite{L2001}, while that of ${\sf RC}$ is large in general.
From these facts, the shifting may possibly be explained by how the variance of ${\sf LRC}(D)$ changes with $m$ and $D$. We leave the further analysis as a future problem.

\section{On the basis dependence of a RC } \label{Fixed_basis}

\begin{table}[tb!] 
  \centering
  \caption{The parameters used to generate the data sets for ${\sf RC}$ without pre-processing and ${\sf Haar}$. }
  \begin{tabular}{c c}
    \hline
    Explanation & Value \\
    \hline \hline
    \# of training data  & $O_{train} = 60000$ \\
    \# of valid data & $O_{valid} = 20000$ \\
    %\# of test data $O_{test}$ & 20000 \\
    \# of sampled unitaries & $N_u = 1000$ \\
    \# of measurements shots & $N_s = 1000$ \\
    \# of qubits & $N_q = 4$ \\
    $k$-bit correlations & $k = 4$\\
    Order of the moment & $k' = 4$\\
    \hline
    \label{table_paras_teacherdata_fixed0}
  \end{tabular}
\end{table}

We here provide the classification results obtained without the pre-processing, i.e., $P = I $ in Eq.~\eqref{eq of quantum circuit}. 
To this end, we first construct classifiers from the data set generated by ${\sf RC}$ without pre-processing and that by ${\sf Haar}$. See TABLE~\ref{table_paras_teacherdata_fixed0} for the parameters in the data set.
We use various learning algorithms, LR, LSVM, SVM, RF and NN.
For each algorithm, we construct 10 classifiers by shuffling the data sets, as explained in Subsec.~\ref{SS:TM}.
The success probability of classification, averaged over 10 classifiers, is given in TABLE~\ref{table_asp_4q_fixed0}, where we clearly observe that the classifiers constructed by all algorithms succeeds to distinguish ${\sf RC}$ and ${\sf Haar}$ with excellent accuracy.

\setlength{\tabcolsep}{6pt} % Default value: 6pt
\renewcommand{\arraystretch}{1.2} % Default value: 
\begin{table}[tb!]
  \centering
  \caption{Average success probability of classification of the training, validation and test data set by various machine learning algorithms. }
  \begin{tabular}{c c c c}
    \hline
    Algorithms & training & validation & test \\
    \hline  \hline
    LR   & 1.000$\pm$0.000 & 1.000$\pm$0.000 & 1.000$\pm$0.000 \\
    LSVM & 1.000$\pm$0.000 & 1.000$\pm$0.000 & 1.000$\pm$0.000 \\
    SVM  & 1.000$\pm$0.000 & 1.000$\pm$0.000 & 1.000$\pm$0.000 \\
    RF   & 1.000$\pm$0.000 & 1.000$\pm$0.000 & 1.000$\pm$0.000 \\
    NN   & 0.966$\pm$0.101 & 1.000$\pm$0.000 & 1.000$\pm$0.000 \\
    \hline
    \label{table_asp_4q_fixed0}
  \end{tabular}
\end{table}

\begin{figure}[tb!]
  \centering
  \includegraphics[width=8cm]{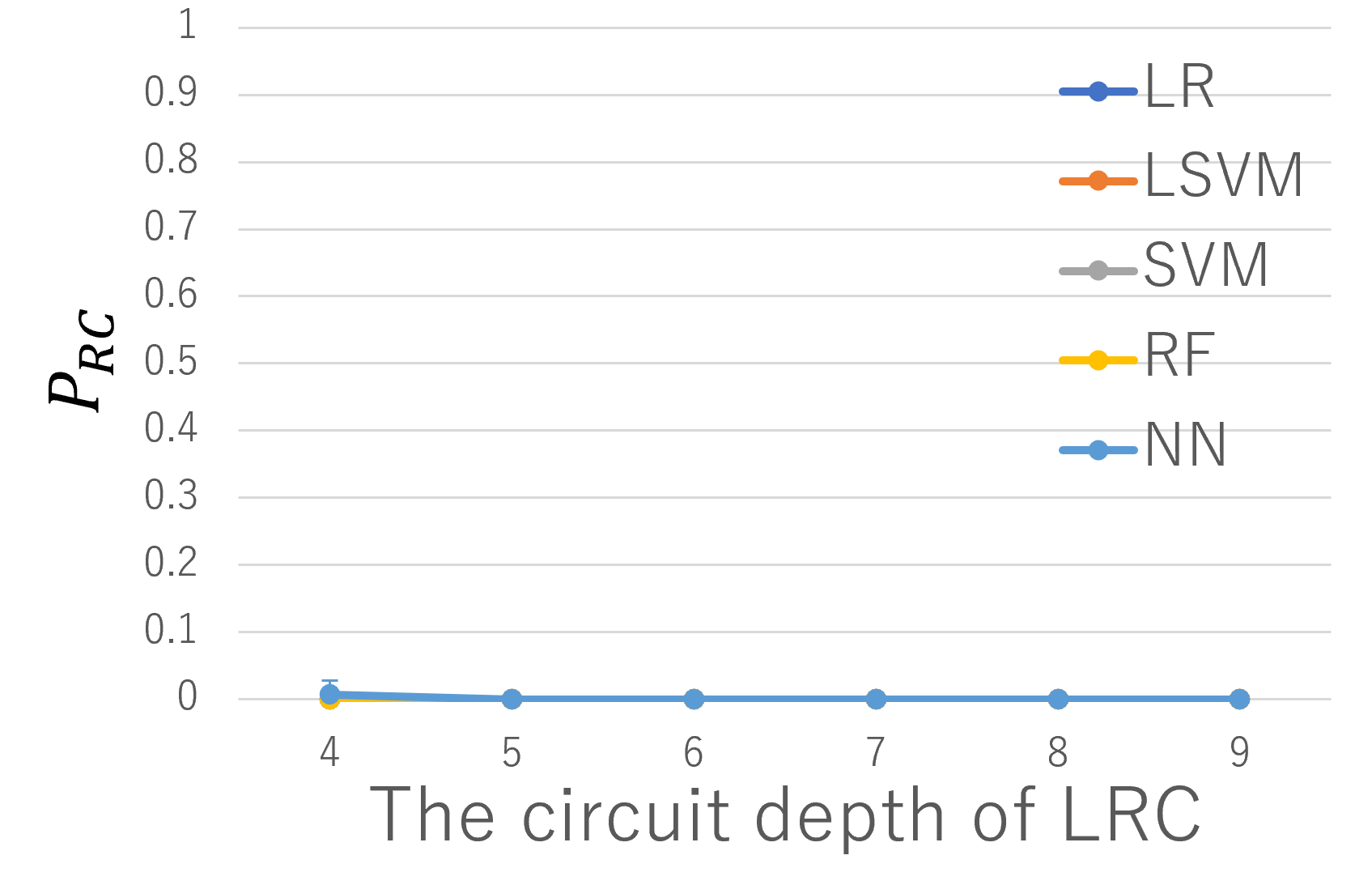}
  \caption{Classification results for the data set generated by ${\sf LRC}(D)$ when the classifiers described in TABLE~\ref{table_asp_4q_fixed0} are applied.}
  \label{pic_LRC_fixed0}
\end{figure}

We then apply the classifiers to the data sets generated by ${\sf LRC}(D)$ for various depth. The classification results are given in FIG.~\ref{pic_LRC_fixed0} and clearly fail to characterize ${\sf LRC}(D)$.
This is likely because Clifford unitaries can generate only a specific class of states, known as the stabilizer states, if they are applied to a computational-basis state. Since stabilizer states have many distinct features, such as that the measurement probability in the computational basis is $2^{-a}$, where $a = 0, \dots, N_q$, the classifiers have learnt such properties specific to stabilizer states.

\section{Parameters used in supervised machine learning} \label{ML_algos}
\begin{figure}[tb!]
  \centering
  \includegraphics[width=9cm]{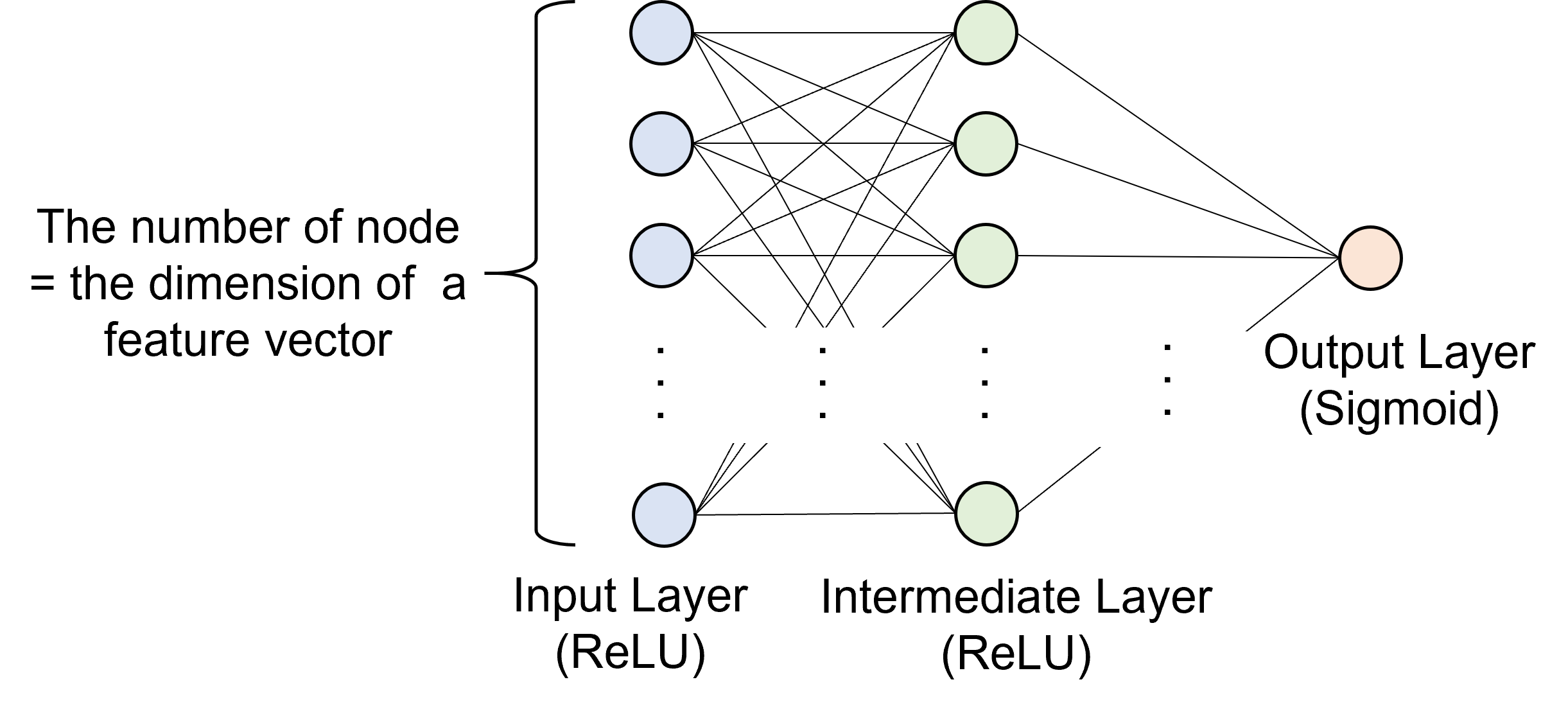}
  \caption{This is the image of our NN model.}
  \label{pic_nn_model}
\end{figure}

We use five machine learning algorithms, linear regression (LR), linear support vector machine (LSVM), support vector machine (SVM), random forest (RF), and neural network (NN). Here, we summarize numerical settings and parameters. For LR, LSVM, SVM, and RF, we use scikit-learn~\cite{scikit-learn} for numerical analysis. While we tried several non-default settings, we eventually chose hyperparameters as the default settings provided by scikit-learn. The exception is that we change the "dual" flag of the LSVM model from True to False, which makes LSVM solve a primal optimization problem.
%In LR, we use the linear regression model implemented in scikit-learn~\cite{scikit-learn}. The regression model is represented by a linear combination of the number of dimensions of the vector of a feature vector, and the value of the regression coefficient is determined so that the residuals of the training data with the model become small. To determine the regression coefficients, the value in the case of the Haar measure is set to $0$ and $1$ in the case of RCs. To discriminate the test data with this model, we calculate the average success probability of discrimination and the percentage discriminated as RC so that if the regression result of the test data was less than 0.5, the model discriminated as the Haar measure, and if it was greater than 0.5, the model discriminated as RCs.\par
%About LSVM, SVM and RF, we use directly from scikit-learn's implementation, but a parameter of LSVM is changed. In implementation of LSVM by scikit-learn, there is a parameter "dual" which selects the algorithm to either solve the dual or primal optimization problem. Default is True, but we set False to solve the priminal optimization problem. The parameters of SVM and RF are default. In SVM, there is a parameter "kernel" which we can select the algorithm to solve the dual of optimization problem by kernel trick. We can choose $4$ pre-implemented kernel, "linear", "poly", "rbf"(default) and "sigmoid". We test all kernels and get hte result that only rbf kernel can discriminate based on the order of $t$-design, so that it is same as default. The version of scikit-learn is 0.23.1.\par
For the NN model, we use Keras~\cite{chollet2015keras} and Tensorflow~\cite{tensorflow2015-whitepaper}. The structure of our NN model is shown in Fig.\,\ref{pic_nn_model}.
We use a fully-connected three-layer neural network. We choose the ReLU and Sigmoid functions as the activation functions for the first two and last layers. We use a binary cross-entropy as a loss function, and the neural network is trained with the Adam optimizer. We set the learning rate of Adam as $10^{-5}$, batch size as $128$, and epoch as $100$.
Note that while we increased the number of hidden layers and units per layer, the performance is not improved. 
%It has $3$ layers, consisted of one intermediate layer sandwiched between an input layer and an output layer. In every node in each layer are fully-connected, and the number of node in the intermediate layer is same as the number of node in the input layer, that is same as the dimension of a feature vector. the number of node in the output layer is one. We choose the ReLU function as an activation function in the input and intermediate layer, (Logistic) Sigmoid function in the output layer. As a optimization of edges, we select Adam algorithm, and we set (Binary) Cross Entropy as a loss function. About the hyperparameters, we set the learning rate of Adam as $1.0\times10^{-5}$, batch size is $128$ and epoch is $100$. A preserving model after training is the model which has the edge parameters that we got the highest average success probability of discrimination for validation dataset.\par
%We have tested some pattern of hyperparamters like increasing the number of an intermediate layers and node, adding special layers calculating BatchNormalization, but they does not exceed the result which the setting is described above.

\section{Computing environments} \label{append_envs}
We used four computers for numerical evaluation. The specifications of the computers are listed in TABLE~\ref{table_hardware_spec}.
\begin{table*}[t]
  \centering
  \caption{The list of computer specifications. We use all computers to generate learning and test data, and use the No.4 to run supervised learning.}
  \begin{tabular}{c c c c}
    \hline
    No & CPU & CPU Core(Thread) & GPU \\
    \hline \hline
    1 & Intel\textregistered Xeon\textregistered Platinum 8276 @2.20GHz & 28(28)$\times$2 & - \\
    2 & Intel\textregistered Xeon\textregistered E5-2695 v4 @2.10GHz $\times$2 & 18(36)$\times$2 & - \\
    3 & Intel\textregistered Core\texttrademark i9-10900K @3.70GHz & 10(20) & - \\
    4 & Intel\textregistered Core\texttrademark i7-8700 @3.20GHz & 6(12) & GeForce GTX-1080 \\
    \hline
    \label{table_hardware_spec}
  \end{tabular}
\end{table*}
The software and its versions used in this paper are summarized in TABLE~\ref{table_software_ver}. Note that we used the dev-branch of qulacs since the version 0.3.0 had a memory leak issue.
\begin{table}[tb!]
  \centering
  \caption{Table of versions of software we mainly used in the computer No.4. It has two virtual environments for running a program with CPU only or CPU and GPU. Below "CPU env" and "GPU env" are corresponding to that environments respectively.}
  \begin{tabular}{c c c}
    \hline
    Software & CPU env & GPU env \\
    \hline  \hline
    python & 3.8.11 & 3.8.11 \\
    tensorflow & 2.4.3 & 2.4.0 \\
    numpy & 1.21.2 & 1.19.5 \\
    scipy & 1.7.1 & 1.6.2 \\
    scikit-learn & 0.24.2 & 0.24.2 \\
    qulacs & 0.3.0("dev" branch) & - \\
    MSVC(C-compiler) & 2017 & - \\
    \hline
    \label{table_software_ver}
  \end{tabular}
\end{table}

We measured the execution times for generating training data and running learning algorithm. The execution times are listed in TABLE~\ref{table_elapsed_time}. 
\begin{table*}[tb!]
  \caption{Table of total running time for data generation. "Computer No" corresponds to the "No" in TABLE~\ref{table_hardware_spec}. "Size" is the size of the datasets. As for the "Time" column, we count the time for generating training and test datasets with $k'$th moments of $k$-bit correlations for all $k\in \{1,\cdots, N_q\}$ and $k'\in \{1,\cdots,20\}$. "RM-LRC" and "N-LRC" in the column "Circuit" means the circuits for noisy LRC and monitored LRC, respectively.}\begin{tabular}{c c c c c c c c}
    \hline
    Computer No & Circuit & depth & Size & $N_u$ & $N_s$ & $N_q$ & time \\
    \hline \hline
    1 & RC & - & 10000 & 20000 & 10000 & 10 & 3.0 days \\
    2 & Haar & - & 10000 & 20000 & 10000 & 10 & 4.0 days \\
    2 & LRC & 20 & 1000 & 20000 & 10000 & 10 & 9.5 hours \\
    2 & RM-LRC & 100 & 1000 & 20000 & 10000 & 10 & 16.7 hours \\
    2 & Haar & - & 10000 & 10000 & 10000 & 7 & 5.0 hours \\
    2 & LRC & 20 & 1000 & 10000 & 10000 & 7 & 30.0 minutes \\
    2 & N-LRC & 15 & 1000 & 10000 & 10000 & 7 & 14.3 hours \\
    2 & RM-LRC & 15 & 1000 & 10000 & 10000 & 7 & 34.0 minutes \\
    2 & RDC & 5 & 1000 & 10000 & 10000 & 7 & 33.5 minutes \\
    3 & RC & - & 10000 & 10000 & 10000 & 7 & 3.0 days \\
    \hline
    \label{table_elapsed_time}
  \end{tabular}
\end{table*}
It takes a long time to generate datasets for large quantum circuits. In contrast, the time taken for learning datasets is much shorter than that. It takes about $5$ minutes to train the NN model with and without GPU. Other learning algorithms take a shorter time than NN, and it finishes within up to a few minutes.

\end{document}